\documentclass[aps,prd,preprint,superscriptaddress,amsmath,amssymb,showpacs,nofootinbib]{revtex4-1}
\usepackage{dcolumn}
\usepackage{graphicx}
\usepackage{float}
\usepackage{physics}
\usepackage[colorlinks=true,allcolors=blue]{hyperref}
\usepackage{mathrsfs}
\usepackage{inputenc}
\usepackage{appendix}
\usepackage[T1]{fontenc}
\usepackage{babel}
\usepackage{geometry}
\begin{document}

\title{Axial Chiral Vortical Effect in a Sphere with finite size effect}
\author{Shu-Yun Yang}
\thanks{yangsy@mails.ccnu.edu.cn}
\affiliation{ Institute of Particle Physics and Key Laboratory of Quark and Lepton Physics (MOS), Central China Normal University, Wuhan 430079, China}

\author{Ren-Hong Fang}
\thanks{fangrh@sdu.edu.cn}
\affiliation{Key Laboratory of Particle Physics and Particle Irradiation (MOE), Institute of Frontier and Interdisciplinary Science, Shandong University, Qingdao, Shandong 266237, China}

\author{De-Fu Hou}
\thanks{Co-corresponding author: houdf@mail.ccnu.edu.cn}
\affiliation{ Institute of Particle Physics and Key Laboratory of Quark and Lepton Physics (MOS), Central China Normal University, Wuhan 430079, China}

\author{Hai-Cang Ren}
\thanks{Co-corresponding author: renhc@mail.ccnu.edu.cn}
\affiliation{ Physics Department, The Rockefeller University, 1230 York Avenue, New York, NY 10021-6399}
\affiliation{ Institute of Particle Physics and Key Laboratory of Quark and Lepton Physics (MOS), Central China Normal University, Wuhan 430079, China}

\begin{abstract}
 We investigate the axial vortical effect in a uniformly rotating sphere subject to finite size. We use MIT boundary condition to limit the boundary of the sphere. For massless fermions inside the sphere, we obtain the exact axial vector current far from the boundary that matches the expression obtained in cylindrical coordinates in the literature. On the spherical boundary, we find both the longitudinal and transverse(with respect to the rotation axis) components with magnitude depending on the colatitude angle. For massive fermions, we derive an expansion of the axial conductivity far from the boundary to all orders of mass whose leading order term agrees with the mass correction reported in the literature. We also obtain the leading order mass correction on the boundary which is linear, and stronger than the quadratic dependence far from the boundary. The qualitative implications on the phenomenology of heavy ion collisions are speculated.
\end{abstract}

\maketitle
\section{Introduction}
	 	   	 	    Relativistic heavy ion collisions (RHIC) are utilized to produce quark-gluon plasmas (QGP) at high temperature and nonzero baryon density. A typical (off-central) collision exposes the QGP thus generated under an ultra-strong magnetic field and endows it with high angular momentum. A number of novel transport phenomena~\cite{Vilenkin:1980fu,Kharzeev:2004ey,Kharzeev:2010gr,Son:2012wh,Fukushima:2008xe,Zhang:2020ben, Son:2009tf,Neiman:2010zi,Lin:2018aon, Golkar:2012kb,Chernodub:2016kxh,Shitade:2020lfe,Abramchuk:2018jhd,Hou:2012xg,Flachi:2017vlp,Lin:2018aon} have been proposed, among them is the Axial-Chiral-Vortical-Effect (ACVE). ACVE refers to the axial vector current, i.e., the spin density of fermions in response to the global angular momentum, and is expected to be detected via the polarization of Lambda post hadronization. ACVE is also expected inside the core of a fast spinning neutron star~\cite{Endrizzi:2018uwl,Lonardoni:2019ypg,Grams:2021qpj}. In this work, we shall focus our attention on the theoretical aspect of ACVE.

In a thermal equilibrium ensemble, the ACVE is represented in terms of the global angular velocity $\omega$ by the formula
\begin{equation}
\boldsymbol{J}_A = \sigma\boldsymbol{\omega}+... ,
\label{J_5}
\end{equation}
where the coefficient $\sigma$ is referred to as the axial vortical conductivity and the ellipsis represents higher power in $\omega$. Through a pioneer work by Son and Surowka ~\cite{Son:2009tf,Pu:2012wn} and a supplemental work by Neiman and Oz~\cite{Neiman:2010zi}, the axial vortical conductivity is restricted by thermodynamic laws to the following general form in the chiral limit,
\begin{equation}
\sigma = \frac{\mu_V^2+\mu_A^2}{2\pi^2} + cT^2  ,
\label{AVE_general}
\end{equation}
where $\mu_V$ and $\mu_A$ are the vector and axial vector chemical potentials, and the coefficient $c$ in front of the temperature square has to be determined by other means.

Besides the hydrodynamic approach,  Eq.~(\ref{AVE_general}) with $c=1/6$ was first derived by Vilenkin via the solution of a free Dirac equation in a rotating cylinder \cite{Vilenkin:1979ui,Vilenkin:1980zv,Vilenkin:1978hb}, and the axial vortical conductivity for non-interacting fermions reads
\begin{equation}
\sigma = \frac{\mu_V^2+\mu_A^2}{2\pi^2} + \frac{1}{6}T^2  .
\label{AVE_free}
\end{equation}
The same expression was obtained by Landsteiner et. al. via the Kubo formula to one-loop order~\cite{Landsteiner:2011cp}. There have been also a large body of literature on the derivation of  Eq.~(\ref{AVE_free}) from kinetic theory~\cite{Gao:2012ix,Yang:2020mtz} or holography~\cite{Torabian:2009qk,Rebhan:2009vc}. Beyond  Eq.~(\ref{AVE_free}), the authors of~\cite{Hou:2012xg, Golkar:2012kb} discovered higher order corrections to the coefficient $c$ in QED or QCD coupling, and the authors of~\cite{Ambrus:2014uqa} figured out the higher order terms in $\omega$, i.e. the ellipsis in  Eq.~(\ref{J_5}) for massless fermions and ended up a closed form of the axial-vector current and the authors of~\cite{Lin:2018aon} derived the leading order correction of the fermion mass. A recent calculation~\cite{Palermo:2021hlf} of axial current for massless fermions in a general thermodynamic equilibrium with rotation and acceleration (within a formalism “far from the boundary”, that is without enforcing boundary conditions) reproduces
the known results for rotating equilibrium, such as in Ref.~\cite{Ambrus:2014uqa}, but it extends to systems including acceleration.

In this work, we explore the axial vortical effect in a finite sphere of radius $R$ subject to the MIT boundary condition. Our motivation is twofold: Firstly, a system rotating with constant angular velocity has to be finite in the direction transverse to the rotation axis as restricted by the subluminal linear speed on the boundary. Secondly, a finite sphere serves as a better approximation to the shape of the QGP fireball in heavy ion collisions and the quark matter core of a neutron star than an infinitely long cylinder considered in literature. The MIT boundary condition effectively separates the deconfinement phase of the interior and the confinement phase outside. But we were unable to include the strong coupling underlying the near-perfect fluidity inside an actual QGP fireball, nor to describe its rapid expansion, especially in the early stage of its evolution. Far from the boundary where the finite size effect can be ignored, we reproduce in spherical coordinates exactly the same form of the axial-vector current in chiral limit derived in cylindrical coordinates~\cite{Ambrus:2014uqa}. We also carry out the fermion mass correction to all orders with the leading order matching the result in Ref.~\cite{Lin:2018aon} which was derived with the Kubo formulation. The infinite series in powers of the mass correction indicates that the leading order correction for the mass of $s$ quark at the RIHC temperature is quite accurate. More importantly, we figure out an analytic approximation of the axial vector current on the spherical boundary with the aid of the asymptotic formula of the Bessel function of large argument and large order. For $\boldsymbol{\omega}=\omega\hat{\boldsymbol{z}}$, we find that
\begin{equation}
\boldsymbol{J}_A=(\sigma\hat{\boldsymbol{z}}+\sigma^\prime\boldsymbol{e}_\rho)\omega  ,
\label{AVEvector}
\end{equation}
with $\boldsymbol{e}_\rho$ the unit radial vector of the cylindrical coordinate systems $(\rho, \phi, z)$. For $T\gg1/R$ and the fermion mass $M\ll T$, the axial vortical conductivity parallel $\boldsymbol{\omega}$ is
\begin{equation}
\sigma = \left\{\frac{\mu^2}{2\pi^2} + \frac{1}{6}T^2-\frac{M}{4\pi}\left[\mu+2T\ln \left(1+e^{-\frac{\mu}{T}}\right)\right]\right\}\cos^2\theta  ,
\label{anisotropicity1}
\end{equation}
and that perpendicular to $\boldsymbol{\omega}$ is
\begin{equation}
\sigma^\prime = \left\{\frac{\mu^2}{16\pi^2} + \frac{1}{48}T^2-\frac{M}{32\pi}\left[\mu+2T\ln \left(1+e^{-\frac{\mu}{T}}\right)\right]\right\}\sin2\theta  ,
\label{anisotropicity2}
\end{equation}
with $\theta$ the polar angle with respect to the direction of the angular velocity. Notice that we have to set $\mu_A=0$ and $\mu_V=\mu$ because the MIT boundary condition breaks the chiral symmetry even for massless fermions. To our knowledge, the perpendicular component has never been reported in the literature and its existence may shed some light on the longitudinal (with respect to the beam direction) polarization in heavy ion collisions.

The organization of the paper is as follows. In Sec.~\ref{sec:12}, general properties of the axial vortical effect are discussed from symmetry perspectives. In Sec.~\ref{sec:2}, we lay out the general formulation of the chiral magnetic effect in spherical coordinates with the MIT boundary condition. The axial vortical effect of massless and massive fermions are calculated in Sec.~\ref{sec:3} and Sec.~\ref{sec:4}. Sec.~\ref{sec:5} concludes the paper with a qualitative speculation on the impact of the finite size effect for heavy ion collisions. Some technical details are deferred to Appendices. We also include two additional Appendices for self-containdness, one for an alternative derivation of the closed end formula of the axial-current in cylindrical coordinates and the other one for the mass correction via the Kubo formula under dimensional regularization. Throughout the paper, we shall stay with the notation of  Eqs.~(\ref{anisotropicity1}) and (\ref{anisotropicity2}) by setting $\mu_A=0$ and $\mu\equiv\mu_V$. Furthermore, the size of the sphere is assumed sufficiently large in comparison with the length scale corresponding to the temperature or chemical potential for the boundary condition to be analytically soluble.

\section{Symmetry consideration}\label{sec:12}

In this section, we explore the axial vortical effect from symmetry perspectives. The validity of the conclusion reached here is not mostly limited to a free Dirac considered in the literature and the subsequent sections of this work.

The axial vortical effect refers to the thermal average of the spatial component of the axial vector current density $\boldsymbol{\mathcal{J}}_A$ in the presence of a nonzero angular momentum. Taking the direction of the angular momentum as $z$-axis, we have
\begin{equation}
\langle \boldsymbol{\mathcal{J}}_{A}(\boldsymbol{r})\rangle=\rm{Tr}\varrho(\mu,\omega)\boldsymbol{\mathcal{J}}_{A}(\boldsymbol{r})\equiv \boldsymbol{J}_{A}(\boldsymbol{r}).
\label{ensemble}
\end{equation}
In terms of the field theoretic Hamiltonian $\mathcal{H}$, conserved charge $\mathcal{Q}$ and $z$-component of the angular momentum $\mathcal{J}_{z}$, the density matrix at thermal equilibrium reads
\begin{equation}
\varrho(\mu,\omega)=Z^{-1}\exp\left(\frac{\mathcal{H}-\mu \mathcal{Q}-\omega \mathcal{J}_z}{T}\right),
\label{varrho}
\end{equation}
where $T$ is the temperature, $\mu$ is the chemical potential, $\omega$ is the angular velocity and $Z$ is the normalization constant such that ${\rm Tr}\varrho=1$.

Introducing the basic vector of cylindrical coordinates $\hat{\boldsymbol{z}}$ and
\begin{eqnarray}
\boldsymbol{e}_\rho(\phi)&=&\hat{\boldsymbol{x}}\cos\phi+\hat{\boldsymbol{y}}\sin\phi\nonumber  ,\\
\boldsymbol{e}_\phi(\phi)&=&-\hat{\boldsymbol{x}}\sin\phi+\hat{\boldsymbol{y}}\cos\phi  ,
\end{eqnarray}
the ensemble average (\ref{ensemble}) can be decomposed into its longitudinal component
\begin{equation}
J_{A}^z(\rho,\phi,z|\mu,\omega)=\hat{\boldsymbol{z}}\cdot\boldsymbol{J}_{A}(\rho,\phi,z|\mu,\omega),
\label{longitudinal}
\end{equation}
and its transverse components
\begin{equation}
J_{A}^\pm(\rho,\phi,z|\mu,\omega)=\boldsymbol{e}_\pm(\phi)\cdot\boldsymbol{J}_{A}(\rho,\phi,z|\mu,\omega)  ,
\label{transverse}
\end{equation}
with
\begin{equation}
\boldsymbol{e}_\pm(\phi)=\frac{1}{\sqrt{2}}(\boldsymbol{e}_\rho\pm i\boldsymbol{e}_\phi) ,
\end{equation}
where the dependence on the cylindrical coordinates, chemical potential, and angular velocity is explicitly indicated and will be suppressed in subsequent sections. We have
\begin{equation}
J_{A}^-(\rho,\phi,z|\mu,\omega)=J_{A}^{+*}(\rho,\phi,z|\mu,\omega) ,
\end{equation}
and consequently
\begin{eqnarray}
J_{A}^\rho(\rho,\phi,z|\mu,\omega) &=& \boldsymbol{e}_\rho(\phi)\cdot\boldsymbol{J}_{A}(\rho,\phi,z|\mu,\omega)=\sqrt{2} {\rm Re}J_{A}^+(\rho,\phi,z|\mu,\omega) ,\\
J_{A}^\phi(\rho,\phi,z|\mu,\omega) &=& \boldsymbol{e}_\phi(\phi)\cdot\boldsymbol{J}_{A}(\rho,\phi,z|\mu,\omega)=\sqrt{2} {\rm Im}J_{A}^+(\rho,\phi,z|\mu,\omega)\nonumber  .
\end{eqnarray}

Assuming that the Hamiltonian and the boundary condition are invariant under spatial rotation, spatial inversion, time reversal, and charge conjugation, we have
\begin{equation}
\mathcal{R(\alpha)}\varrho(\mu,\omega)\mathcal{R(\alpha)}^{-1}=\varrho(\mu,\omega)
\label{rotation0}   ,
\end{equation}
\begin{equation}
\mathcal{P}\varrho(\mu,\omega)\mathcal{P}^{-1}=\varrho(\mu,\omega)   ,
\label{parity0}
\end{equation}
\begin{equation}
\mathcal{T}\varrho(\mu,\omega)\mathcal{T}^{-1}=\varrho(\mu,-\omega)   ,
\end{equation}
\begin{equation}
\mathcal{C}\varrho(\mu,\omega)\mathcal{C}^{-1}=\varrho(-\mu,\omega)   ,
\label{Charge}
\end{equation}
where $\mathcal{R}(\alpha)$ is a Hilbert space operator of a rotation about $z$-axis by an angle $\alpha$, and $\mathcal{P}$, $\mathcal{T}$ and $\mathcal{C}$ are Hilbert space operators for spatial inversion, time reversal, and charge conjugation. Together with transformation laws of the axial vector current $\boldsymbol{\mathcal{J}}_{A}(\boldsymbol{r})$
\begin{equation}
\mathcal{R(\alpha)}\boldsymbol{\mathcal{J}}_{A}(\rho,\phi,z)\mathcal{R(\alpha)}^{-1}=\overleftrightarrow{D}(\alpha)\cdot\boldsymbol{\mathcal{J}}_{A}(\rho,\phi-\alpha,z) ,
\end{equation}
\begin{equation}
\mathcal{P}\boldsymbol{\mathcal{J}}_{A}(\rho,\phi,z)\mathcal{P}^{-1}=\boldsymbol{\mathcal{J}}_{A}(\rho,\phi+\pi,-z)  ,
\end{equation}
\begin{equation}
\mathcal{T}\boldsymbol{\mathcal{J}}_{A}(\rho,\phi,z)\mathcal{T}^{-1}=-\boldsymbol{\mathcal{J}}_{A}(\rho,\phi,z)  ,
\end{equation}
\begin{equation}
\mathcal{C}\boldsymbol{\mathcal{J}}_{A}(\rho,\phi,z)\mathcal{C}^{-1}=\boldsymbol{\mathcal{J}}_{A}(\rho,\phi,z)  ,
\end{equation}
it follows that
\begin{equation}
\boldsymbol{J}_{A}(\rho,\phi,z|\mu,\omega)={\rm Tr}\mathcal{R}(\alpha)\varrho(\mu,\omega)\boldsymbol{\mathcal{J}}_{A}(\rho,\phi,z)\mathcal{R}^{-1}(\alpha)
=\overleftrightarrow{D}(\alpha)\cdot\boldsymbol{J}_{A}(\rho,\phi-\alpha,z|\mu,\omega)  ,
\label{rotation}
\end{equation}
\begin{equation}
\boldsymbol{J}_{A}(\rho,\phi,z|\mu,\omega)={\rm Tr}\mathcal{P}\varrho(\mu,\omega)\boldsymbol{\mathcal{J}}_{A}(\rho,\phi,z)\mathcal{P}^{-1}=\boldsymbol{J}_{A}(\rho,\phi+\pi,-z|\mu,\omega)  ,
\label{parity}
\end{equation}
\begin{equation}
\boldsymbol{J}_{A}(\rho,\phi,z|\mu,\omega)={\rm Tr}\mathcal{T}\varrho(\mu,\omega)\boldsymbol{\mathcal{J}}_{A}(\rho,\phi,z)\mathcal{T}^{-1}=-\boldsymbol{J}_{A}(\rho,\phi,z|\mu,-\omega) ,
\label{reversal}
\end{equation}
\begin{equation}
\boldsymbol{J}_{A}(\rho,\phi,z|\mu,\omega)={\rm Tr}\mathcal{C}\varrho(\mu,\omega)\boldsymbol{\mathcal{J}}_{A}(\rho,\phi,z)\mathcal{C}^{-1}=\boldsymbol{J}_{A}(\rho,\phi,z|-\mu,\omega) ,
\label{conjugation}
\end{equation}
where $\overleftrightarrow{D}(\alpha)$ is the dyadic notation of the $3\times 3$ rotation matrix
\begin{equation}
\overleftrightarrow{D}(\alpha)=\left(\begin{array}{ccc}
\cos\alpha & -\sin\alpha & 0\\
\sin\alpha & \cos\alpha & 0\\
0 & 0 & 1
\end{array}\right).
\end{equation}
Because of the relations
\begin{equation}
\boldsymbol{e}_s(\phi)\cdot\overleftrightarrow{D}(\alpha)=\boldsymbol{e}_s(\phi-\alpha),(s=\pm1)
\end{equation}
and $\boldsymbol{e}_s(\phi+\pi)=-\boldsymbol{e}_s(\phi)$, Eq.~(\ref{rotation}) and Eq.~(\ref{parity}) imply that longitudinal and transverse components of the axial current defined in Eq.~(\ref{longitudinal}) and Eq.~(\ref{transverse}) are independent of the azimuthal angle as expected and the transverse component is odd in $z$, i.e.
\begin{equation}
J_{A}^s(\rho,\phi,-z|\mu,\omega)=-J_{A}^s(\rho,\phi,z|\mu,\omega) .
\label{symmetry}
\end{equation}
Consequently, there cannot be a transverse axial vortical effect for an infinitely long cylinder since the axial vector current is independent of $z$. This, however, is not the case with a sphere
as the $z$ dependence cannot be ignored. The oddness with respect to $z$ implies only zero transverse axial vector current on the equatorial plane of the sphere. Indeed, the subsequent sections show that the transverse component of the axial-vector current does exist on the spherical boundary for a free Dirac field and does vanish on the equatorial plane. The other two Eqs. (\ref{reversal}) and (\ref{conjugation}) imply that the thermal average of the axial-vector current is always odd with respect to the angular velocity and even with the chemical potential.

Before concluding this section, we remark that some of the relations above can be readily generalized to a non-equilibrium density matrix with its time development dictated by the Liouville theorem. For instance \footnote{The original MIT boundary condition in ~\cite{Chodos:1974je} also applies to a time-dependent shape of a bag}, for a homogeneous and expanding system, as long as relations (\ref{rotation0}), (\ref{parity0}) and (\ref{Charge}) hold initially, they will hold for all time. Then relation (\ref{symmetry}) and its implications discussed above remain valid for all time.

\section{Axial vector current in spherical coordinates}\label{sec:2}
\subsection{Hamiltonian}\label{sec:2a}
The Hamiltonian for a Dirac fermion in a uniformly rotating
system with angular velocity $\boldsymbol{\omega}=\omega\boldsymbol{e}_{z}$
can be written as~\cite{Ambrus:2014uqa,Chen:2019tcp}
\begin{equation}
H=H_{0}-\omega J_{z}-\mu,\label{eq:pq1}
\end{equation}
where $H_{0}=-i\boldsymbol{\alpha}\cdot\boldsymbol{\nabla}+\beta M$
is the free Hamiltonian, $J_{z}=\frac{1}{2}\Sigma_{3}-i(x\partial_{y}-y\partial_{x})$
is the $z$-component of the total angular momentum, $\mu$ is the
the chemical potential of the system, $M$ is the mass of the Dirac fermion,
and $\boldsymbol{\alpha}=\gamma^{0}\boldsymbol{\gamma}$, $\beta=\gamma^{0}$.
We work in the Dirac representation for gamma matrices $\gamma^{\mu}$
as follows,
\begin{equation}
\gamma^{0}=\left(\begin{array}{ccc}
1 & 0\\
0 & -1
\end{array}\right),\ \ \ \gamma^{i}=\left(\begin{array}{cc}
0 & \sigma^{i}\\
-\sigma^{i} & 0
\end{array}\right).\label{eq:oo1}
\end{equation}
The last two terms of  Eq.~(\ref{eq:pq1}) are included in the single particle Dirac Hamiltonian because it is  Eq.~(\ref{eq:pq1}), when being sandwiched between Dirac fields $\psi, \psi^\dagger$
\begin{equation}
\mathcal{H}=\int d^3\boldsymbol{r}\psi^\dagger(\boldsymbol{r})H\psi(\boldsymbol{r}),
\end{equation}
to define the density operator $e^{(-\frac{\mathcal{H}}{T})}$ for thermal average.

In this section, we consider the eigenfunctions of the Hamiltonian in spherical coordinates. The eigenfunctions of the Hamiltonian satisfy
\begin{equation}
H\psi=(E-\omega J_{z}-\mu)\psi,\label{eq:aa}
\end{equation}
where $E$ is the eigen-energy of $H_{0}$. The solutions of  Eq.~(\ref{eq:aa}) can be chosen as the common eigenfunctions
of these four commutative Hermitian operators: Hamiltonian $H$, square
of total angular momentum $\boldsymbol{J}^{2}$, $z$-component of
total angular momentum $J_{z}$, parity operator $P$. We list the eigenfunctions in spherical coordinates as follows,
\begin{eqnarray}
\psi_{j,l=j+\frac{1}{2},m}(r,\theta,\phi) & = & \left(\begin{array}{c}
f(r)Z_{j,j+\frac{1}{2},m}(\theta,\phi)\\
-ig(r)Z_{j,j-\frac{1}{2},m}(\theta,\phi)
\end{array}\right),\nonumber \\
\psi_{j,l=j-\frac{1}{2},m}(r,\theta,\phi) & = & \left(\begin{array}{c}
f(r)Z_{j,j-\frac{1}{2},m}(\theta,\phi)\\
ig(r)Z_{j,j+\frac{1}{2},m}(\theta,\phi)
\end{array}\right),\label{eq:fa2}
\end{eqnarray}
with $j, l, m$ denoting the eigenvalues $j(j+1), (-1)^l, m$ of $\boldsymbol{J}^{2}, P, J_{z}$
respectively and the spinor spherical harmonics $Z_{j,j\pm\frac{1}{2},m}(\theta,\phi)$ defined as
\begin{eqnarray}
Z_{j,j+\frac{1}{2},m}(\theta,\phi) & = & \frac{1}{\sqrt{2(j+1)}}\left(\begin{array}{c}
\sqrt{j-m+1}Y_{j+\frac{1}{2},m-\frac{1}{2}}(\theta,\phi)\\
-\sqrt{j+m+1}Y_{j+\frac{1}{2},m+\frac{1}{2}}(\theta,\phi)
\end{array}\right),\nonumber \\
Z_{j,j-\frac{1}{2},m}(\theta,\phi) & = & \frac{1}{\sqrt{2j}}\left(\begin{array}{c}
\sqrt{j+m}Y_{j-\frac{1}{2},m-\frac{1}{2}}(\theta,\phi)\\
\sqrt{j-m}Y_{j-\frac{1}{2},m+\frac{1}{2}}(\theta,\phi)
\end{array}\right),\label{eq:fa1}
\end{eqnarray}
 and $f(r)$, $g(r)$ the radial wave functions. Making use of the following relations,
\begin{eqnarray}
\boldsymbol{\sigma\cdot\hat r}Z_{j,j\pm\frac{1}{2},m}(\theta,\phi) &=& Z_{j,j\mp\frac{1}{2},m}(\theta,\phi)\nonumber,\\
\boldsymbol{\sigma\cdot\nabla}\left[f(r)Z_{j,j\pm\frac{1}{2},m}(\theta,\phi)\right] &=& \left[f^{\prime}(r)+\left(1\pm j\pm\frac{1}{2}\right)\frac{f(r)}{r}\right]Z_{j,j\mp\frac{1}{2},m}(\theta,\phi),\label{eq:q3}
\end{eqnarray}
one can obtain following differential equation satisfied by $f(r)$ in  Eq.~(\ref{eq:fa2}),
\begin{equation}
r^{2}f^{\prime\prime}(r)+2rf^{\prime}(r)+\left[r^{2}(E^{2}-M^{2})-l(l+1)\right]f(r)=0,\label{eq:oo2}
\end{equation}
which is the $l$-th order spherical Bessel equation. The radial function $g(r)$ in  Eq.~(\ref{eq:fa2}) can be expressed by $f(r)$,
\begin{equation}
g(r)=\left\{ \begin{array}{cc}
\frac{1}{E+M}\left[f^{\prime}(r)+\frac{l+1}{r}f(r)\right], & \mathrm{for}\ l=j+\frac{1}{2}\\
-\frac{1}{E+M}\left[f^{\prime}(r)-\frac{l}{r}f(r)\right], & \mathrm{for}\ l=j-\frac{1}{2}
\end{array}\right..\label{eq:oo3}
\end{equation}
We list the solutions of $f(r)$ and $g(r)$ in Table~\ref{eq:t1}, with $k>0$, $E_{k}=\sqrt{k^{2}+M^{2}}$ and $C$ the normalization factor.
 \begin{table}[ht]
\caption{Solutions of $f(r)$ and $g(r)$.}
\centering
\begin{tabular}{|c|c|cc|}
\hline $l=j\pm\frac{1}{2}$ &$ E=E_{k} $& $E=-E_{k} $& \\
\hline$ f(r)$ &$ Cj_{l}(kr)$ & $-C\sqrt{\frac{E_{k}-M}{E_{k}+M}}j_{l}(kr)$ & \\
\hline $g(r)$ & $C\sqrt{\frac{E_{k}-M}{E_{k}+M}}j_{l\mp1}(kr)$ & $Cj_{l\mp1}(kr)$ &
\\\hline \end{tabular}  \label{eq:t1}
\end{table}

For a spherical volume of radius $R$, the quantization of the radial momentum $k$ depends on the boundary condition. An approximate boundary condition for a fireball of QGP follows from the MIT bag model~\cite{Chodos:1974je} and reads

\begin{equation}
-i\boldsymbol{\gamma}\cdot\boldsymbol{\hat r}\psi_{j,l,m}|_{|\boldsymbol{r}|=R}=\psi_{j,l,m}|_{|\boldsymbol{r}|=R},\label{eq:fa4}
\end{equation}
which requires that the solution of the Dirac equation on the boundary implements the eigenfunction of $-i\boldsymbol{\gamma}\cdot\boldsymbol{\hat r}$ of eigenvalue one. As $\gamma_5\boldsymbol{\gamma}\cdot\boldsymbol{\hat r}=-\boldsymbol{\gamma}\cdot\boldsymbol{\hat r}\gamma_5$, the MIT boundary condition breaks the chiral symmetry even for massless fermions. It follows from Eq.~(\ref{eq:fa2}) and Eq.~(\ref{eq:q3}) that the radial wave function satisfies
\begin{equation}
f(R)=\pm g(R)  ,
\label{MIT}
\end{equation}
for $l=j\mp 1/2$. For the solutions of the free Dirac equation in Table~\ref{eq:t1} \footnote{For a finite sphere, one has to examine whether there are edge states with $0<E<M$. Setting $\kappa=\sqrt{M^2-E^2}$, the MIT boundary conditions Eq.~(\ref{MIT1}) and Eq.~(\ref{MIT2}) becomes $i_{j-1/2}(\kappa R)=-i_{j+1/2}(\kappa R)\tan\chi^\prime$ and $i_{j+1/2}(\kappa R)=-i_{j-1/2}(\kappa R)\mathrm{\tan}\chi^\prime$ with $i_l$(z) the modified spherical Bessel function and $tan\chi^\prime=\sqrt{\frac{M-E}{M+E}}$. There is no solution for $\kappa$ in either case and the edge states are ruled out by the MIT boundary condition.}, the MIT boundary condition reads
\begin{equation}
j_{j-1/2}(kR)=j_{j+1/2}(kR)\tan\chi  ,
\label{MIT1}
\end{equation}
for the positive energy state of $l=j-1/2$ and
\begin{equation}
j_{j+1/2}(kR)=-j_{j-1/2}(kR)\tan\chi   ,
\label{MIT2}
\end{equation}
for the positive energy state of $l=j+1/2$, where
\begin{equation}
\tan\chi=\sqrt{\frac{E_{k}-M}{E_{k}+M}}.
\label{tangent}
\end{equation}
The boundary conditions for the negative energy states follow from the charge conjugation, i.e.
\begin{equation}
\psi_{j,l,m}^c=\gamma^2\psi_{j,l,m}^* .
\end{equation}
Employing the integration formula
\begin{equation}
\int_0^RdrrJ_\nu^2(kr)=\frac{R^2}{2}\left[J_\nu^{\prime2}(kR)+(1-\frac{\nu^2}{k^2R^2})J_\nu^2(kR)\right] ,
\end{equation}
and the formulas of the derivative $J_\nu^\prime(z)$ in terms of $J_\nu(z)$ and $J_{\nu\pm1}(z)$, the normalization constant in Table~\ref{eq:t1} is readily determined
\begin{equation}
|C|^{2}=\left\{ \begin{array}{cc}
\frac{2}{R^{3}}\left(\sec^{2}\chi+\csc^{2}\chi-\frac{2j}{kR}\cot\chi-\frac{2j+2}{kR}\tan\chi\right)^{-1}j_{j-1/2}^{-2}(kR), & \mathrm{for}\ l=j-\frac{1}{2}\\
\frac{2}{R^{3}}\left(\sec^{2}\chi+\csc^{2}\chi+\frac{2j}{kR}\tan\chi+\frac{2j+2}{kR}\cot\chi\right)^{-1}j_{j+1/2}^{-2}(kR), & \mathrm{for}\ l=j+\frac{1}{2}
\end{array} \right.. \label{eq:norm}
\end{equation}

The boundary conditions Eq.~(\ref{MIT1}) and Eq.~(\ref{MIT2}) can be solved approximately for $kR\gg1$ and $j\ll kR$ with the aid of the asymptotic formula of the spherical Bessel function
\begin{equation}
j_l(x)\simeq\frac{1}{x}\sin\left(x-\frac{l\pi}{2}\right)\ \ \mathrm{as}\ \ x\gg \mathrm{max}(1,l).
\label{eq:asymp}
\end{equation}
We find
\begin{eqnarray}
kR-\frac{l\pi}{2}+\chi & = & n\pi,\ \ (n\in\mathbb{Z})\label{eq:aa6}
\end{eqnarray}
for $l=j\pm1/2$. The summation of $k$ can be converted to an integral
\begin{equation}
\sum_k(...)=\frac{R}{\pi}\int_0^\infty dk(...)  ,
\label{eq:sum_intgl}
\end{equation}
and the normalization constant Eq.~(\ref{eq:norm}) under both conditions Eq.~(\ref{MIT1}) and  Eq.~(\ref{MIT2}) is simplified to
\begin{equation}
|C|=\frac{\sqrt{2}k}{\sqrt{R}}\sqrt{\frac{1}{\tan^{2}\chi+1}}=\frac{k}{\sqrt{R}}\sqrt{\frac{E_{k}+M}{E_{k}}}.
\label{norm}
\end{equation}
\subsection{Quantized Dirac field}
The quantized Dirac field can be expressed by the eigenfunctions of the Hamiltonian $H$ as follows,
\begin{equation}
\psi(\boldsymbol{r})=\sum_{kjlm}\left[a_{kjlm}u_{kjlm}(\boldsymbol{r})+b_{kjlm}^{\dagger}v_{kjlm}(\boldsymbol{r})\right],\label{eq:qd1}
\end{equation}
where $a_{kjlm}^{\dagger}$ and $a_{kjlm}$ are the creation and annihilation operators of particles, $b_{kjlm}^{\dagger}$ and $b_{kjlm}$ are those of anti-particles. The explicit forms of $u_{kjlm}(\boldsymbol{r})$ and $v_{kjlm}(\boldsymbol{r})$ are
\begin{eqnarray}
u_{kjlm}(\boldsymbol{r}) & = & \psi_{jlm}(\boldsymbol{r}),\nonumber \\
v_{kjlm}(\boldsymbol{r}) & = & \gamma^{2}\psi_{jlm}^{*}(\boldsymbol{r}).\label{eq:fa4}
\end{eqnarray}
We have
\begin{eqnarray}
Hu_{kjlm} & = & (E_{k}-m\omega-\mu)u_{kjlm},\nonumber \\
Hv_{kjlm} & = & (-E_{k}+m\omega-\mu)v_{kjlm},\label{eq:fa5}
\end{eqnarray}
where $E_{k}=\sqrt{k^{2}+M^{2}}$. The ensemble average (\ref{ensemble}) of $a_{kjlm}^{\dagger}a_{kjlm}$ and $b_{kjlm}^{\dagger}b_{kjlm}$ with the density operator (\ref{varrho}) gives rise to the Fermi-Dirac distribution functions,
\begin{eqnarray}
\left\langle a_{kjlm}^{\dagger}a_{kjlm}\right\rangle  & = & \frac{1}{e^{\beta(E_{k}-m\omega-\mu)}+1},\nonumber \\
\left\langle b_{kjlm}^{\dagger}b_{kjlm}\right\rangle  & = & \frac{1}{e^{\beta(E_{k}-m\omega+\mu)}+1},\label{eq:fa6}
\end{eqnarray}
and the thermal expectation values of $a_{kjlm}^{\dagger}b_{kjlm}^{\dagger}$,$b_{kjlm}^{\dagger}a_{kjlm}^{\dagger}$, $a_{kjlm}b_{kjlm}$, $b_{kjlm}a_{kjlm}$ are all zero.

In the following, we calculate the axial vector current of the uniformly rotating system of Dirac fermions. The axial vector current is the ensemble average of the corresponding operator, i.e.
\begin{eqnarray}
\boldsymbol{J}_{A} & = & \langle\psi^{\dagger}\boldsymbol{\Sigma}\psi\rangle\nonumber \\
 & = & \boldsymbol{J}_{A}^{\mathrm{vac}}+\sum_{kjlm}\left(\left\langle a_{kjlm}^{\dagger}a_{kjlm}\right\rangle u_{kjlm}^{\dagger}\boldsymbol{\Sigma}u_{kjlm}-\left\langle b_{kjlm}^{\dagger}b_{kjlm}\right\rangle v_{kjlm}^{\dagger}\boldsymbol{\Sigma}v_{kjlm}\right)\nonumber \\
 & = & \sum_{kjlm}\left[\frac{1}{e^{\beta(E_{k}-m\omega-\mu)}+1}+\frac{1}{e^{\beta(E_{k}-m\omega+\mu)}+1}\right]u_{kjlm}^{\dagger}\boldsymbol{\Sigma}u_{kjlm},\label{eq:fa7}
\end{eqnarray}
where $\boldsymbol{J}_{A}^{\mathrm{vac}}=-\sum_{kjlm}v_{kjlm}^{\dagger}\boldsymbol{\Sigma}v_{kjlm}=0$
is the vacuum term, and the charge conjugation relation in  Eq.~(\ref{eq:fa4}) has been employed in the last step. It follows from the relation
\begin{equation}
Z_{j,j\mp\frac{1}{2},m}(\theta,\phi)=\pm i(-)^{m-\frac{1}{2}}Z_{j,j\mp\frac{1}{2},m}^*(\theta,\phi)  ,
\end{equation}
that
\begin{equation}
u_{kjlm}^\dagger\boldsymbol{\Sigma}u_{kjlm}=-u_{kjl-m}^\dagger\boldsymbol{\Sigma}u_{kjl-m},
\end{equation}
and Eq.~(\ref{eq:fa7}) becomes
\begin{equation}
\boldsymbol{J}_{A}=\sum_{kjlm}\left[\frac{1}{e^{\beta(E_{k}-m\omega-\mu)}+1}-\frac{1}{e^{\beta(E_{k}+m\omega+\mu)}+1}\right]u_{kjlm}^{\dagger}\boldsymbol{\Sigma}u_{kjlm}.
\label{eq:fa7_}
\end{equation}

Introducing the following $\phi$ independent functions
\begin{equation}
\zeta_{jlm}(\theta)\equiv Z_{jlm}^\dagger(\theta,\phi)\sigma_3 Z_{jlm}(\theta),\label{j+1}
\end{equation}
and
\begin{equation}
\eta_{jlm}(\theta)\equiv Z_{jlm}^\dagger(\theta,\phi)\sigma_+ Z_{jlm}(\theta,\phi),\label{j+2}
\end{equation}
we have
\begin{eqnarray}
u_{k,j,j\mp\frac{1}{2},m}^\dagger(\boldsymbol{r})\Sigma_3u_{k,j,j\mp\frac{1}{2},m}(\boldsymbol{r})=&|C_\pm|^2\left[j_{j\mp\frac{1}{2}}^2(kr)\zeta_{j,j\mp\frac{1}{2},m}(\theta) \right.\nonumber\\&\left.
+j_{j\pm\frac{1}{2}}^2(kr)\zeta_{j,j\pm\frac{1}{2},m}(\theta)\tan^2\chi_k\right],
\label{j+3}
\end{eqnarray}
and
\begin{eqnarray}
u_{k,j,j\mp\frac{1}{2},m}^\dagger(\boldsymbol{r})\Sigma_+u_{k,j,j\mp\frac{1}{2},m}(\boldsymbol{r})=&|C_\pm|^2\left[j_{j\mp\frac{1}{2}}^2(kr)\eta_{j,j\mp\frac{1}{2},m}(\theta)\right.\nonumber\\&\left.
+j_{j\pm\frac{1}{2}}^2(kr)\eta_{j,j\pm\frac{1}{2},m}(\theta)\tan^2\chi_k \right],\label{j+4}
\end{eqnarray}
with $|C_\pm|^2$ given by the upper (lower) line of Eq.~(\ref{eq:norm})
In particular, the expression of $\eta_{jlm}(\theta)$ can be reduced to
\begin{equation}
\eta_{j,j-\frac{1}{2},m}(\theta)=\frac{1}{2j}e^{-i\phi}Y_{j-\frac{1}{2},m-\frac{1}{2}}^*(\theta,\phi)L_+Y_{j-\frac{1}{2},m-\frac{1}{2}}(\theta,\phi),\label{j+5}
\end{equation}
and
\begin{equation}
\eta_{j,j+\frac{1}{2},m}(\theta)=-\frac{1}{2(j+1)}e^{-i\phi}Y_{j+\frac{1}{2},m-\frac{1}{2}}^*(\theta,\phi)L_+Y_{j+\frac{1}{2},m-\frac{1}{2}}(\theta,\phi),\label{j+6}
\end{equation}
with
\begin{equation}
L_+=e^{i\phi}\left(\frac{\partial}{\partial\theta}+i\cot\theta\frac{\partial}{\partial\phi}\right).
\label{j+7}
\end{equation}
It follows from the property $Y_{lm'}(\pi-\theta,\phi+\pi)=(-)^lY_{lm'}(\theta,\phi)$ that
$\eta_{jlm}(\pi-\theta)=-\eta_{jlm}(\theta)$ and thereby $\eta_{jlm}(\frac{\pi}{2})=0$.

Before concluding this section, we point out an interesting property of the MIT boundary condition, which is not dictated by symmetries:  The axial vector current vanishes along the equator of the fireball. Indeed,  Eq.~(\ref{MIT}) implies that
\begin{equation}
u_{kjlm}^{\dagger}\Sigma_{3}u_{kjlm}=f^2(R)\Theta_{jm}(\theta) ,  \label{eq:qd6}
\end{equation}
with
\begin{equation}
\Theta_{jm}(\theta)=Z_{j,j-\frac{1}{2},m}^\dagger(\theta,\phi)\sigma_3Z_{j,j-\frac{1}{2},m}(\theta,\phi)+Z_{j,j+\frac{1}{2},m}^\dagger(\theta,\phi)\sigma_3Z_{j,j+\frac{1}{2},m}(\theta,\phi).
\label{eq:Theta}
\end{equation}
Writing $\Theta_{jm}(\theta)$ in terms of associated Legendre functions $P_l^\mu(\cos\theta)$ and using the explicit form of $P_l^\mu(0)$, we find that
\begin{equation}
\Theta_{jm}\left(\frac{\pi}{2}\right)=0.
\label{eq:equator}
\end{equation}
See Appendix~\ref{appendix:a} for details of the proof.
\section{axial chiral vortical effect of massless fermions with finite-size effect}\label{sec:3}
\subsection{Axial vector current far from the boundary}
\label{sec:m=0}For massless fermions, $M=0$ and $E_{k}=k$ in Table~\ref{eq:t1}. Far from the boundary, the main support of the axial vector current comes from the spherical Bessel function with $l=O(1)$. Together with the condition $T\gg 1/R$ and $k\sim T$, we have $kR\gg 1$ for typical radial momentum and the approximation in the last paragraph of Sec.~\ref{sec:2a} becomes handy. Using the relations
\begin{equation}
\frac{1}{e^x+1}=1-\frac{1}{e^{-x}+1}.
\label{eq_ren1}
\end{equation}
 the $z$-component of Eq.~(\ref{eq:fa7}) reads
\begin{equation}
J_A^z=\frac{R}{2 \pi} \int_{-\infty}^{\infty} d k \sum_{j l m}\left[\frac{1}{e^{\beta(k-m \omega-\mu)}+1}-\frac{1}{e^{\beta(k+m \omega-\mu)}+1}\right] u_{k j l m}^{\dagger} \Sigma_{3} u_{k j l m},
\label{eq_ren2}
\end{equation}
where we have turned the summation over $k$ to integral according to  Eq.~(\ref{eq:sum_intgl})
and extended the integration domain to $(-\infty,\infty)$ via  Eq.~(\ref{eq_ren1}).

The Taylor expansion of the axial current in  Eq.~(\ref{eq_ren2}) reads
\begin{equation}
J_{A}^{z}=\sum_{n=0}^{\infty}C_{n}\omega^{2n+1},\label{eq:fa8}
\end{equation}
where the coefficient
\begin{equation}
\begin{aligned}C_{n} \equiv &-\frac{R}{(2n+1)!\pi T^{2n+1}}\int_{-\infty}^{\infty}dkf^{(2n+1)}\left(\frac{k-\mu}{T}\right)\sum_{j}\sum_{m=-j}^{j}m^{2n+1}  \\
 &\times\left(u_{k,j,j-\frac{1}{2},m}^{\dagger}\Sigma_{3}u_{k,j,j-\frac{1}{2},m}+u_{k,j,j+\frac{1}{2},m}^{\dagger}\Sigma_{3}u_{k,j,j+\frac{1}{2},m}\right)\\
  =&-\frac{4}{(2n+1)!\pi T^{2n+1}}\int_{-\infty}^{\infty}dkk^{2}f^{(2n+1)}\left(\frac{k-\mu}{T}\right)\sum_{l=0}^{\infty}j_{l}^{2}(kr)\\
  &\sum_{m^{\prime}=-l}^{l}\left(m^{\prime}+\frac{1}{2}\right)^{2n+1}\left|Y_{lm^{\prime}}(\theta,\varphi)\right|^{2}  ,
\end{aligned}
\label{eq_ren3}
\end{equation}
with $f(x)=1/(e^x+1)$ the Fermi-Dirac distribution function and $f^{(n)}(x)$ its $n$-th derivative. In the second step of  Eq.~(\ref{eq_ren3}), we have substituted the explicit form of the wave function in  Eq.~(\ref{eq:qd1}) together with  Eq.~(\ref{eq:fa1}) for the spinor spherical harmonics.  Applying the addition formula
\begin{equation}
\begin{aligned}
j_{0}\left(k\left|\boldsymbol{r}-\boldsymbol{r}^{\prime}\right|\right)=4 \pi \sum_{l=0}^{\infty} j_{0}(k r) j_{0}\left(k r^{\prime}\right) \sum_{m^{\prime}=-l}^{l} Y_{l m^{\prime}}^{*}(\theta, \varphi) Y_{l m^{\prime}}\left(\theta^{\prime}, \varphi^{\prime}\right) ,
\end{aligned} \label{s1}
\end{equation}
for $r^\prime=r$, $\theta^\prime=\theta$ and $\phi^\prime=\phi+\epsilon$, we find
\begin{equation}
C_{n}=\left.i\frac{(-)^{n}}{(2n+1)!\pi^{2} T^{(2n+1)}}\frac{d^{2n+1}}{d\epsilon^{2n+1}}\left[\frac{e^{i\frac{\epsilon}{2}}}{\xi}\int_{-\infty}^{\infty}dkkf^{(2n+1)}\left(\frac{k-\mu}{T}\right)\sin(k\xi)\right]\right|_{\epsilon=0} ,\label{eq:c}
\end{equation}
where $\xi\equiv 2r\sin\theta\sin\frac{\epsilon}{2}$. After $2n$ times of integration by part with respect to $k$, we obtain that
\begin{equation}
C_{n}=\left.\frac{i}{(2 n+1) ! \pi^{2} T^{(2n+1)}} \frac{d^{2 n+1}}{d \epsilon^{2 n+1}}\left\{e^{i \frac{\epsilon}{2}} \xi^{2 n-2} \int_{-\infty}^{\infty} d k k f^{\prime}\left(\frac{k-\mu}{T}\right)[-2 n \cos (k \xi)+k \xi \sin (k \xi)]\right\}\right|_{\epsilon=0}  .\label{eq:c1}
\end{equation}
Only the $(2n+1)$-th power of $\xi$ inside the curly brackets contributes to $C_n$. Together with the integrals
\begin{equation}
\int_{-\infty}^\infty dxf^\prime(x)=-1,  \qquad  \int_{-\infty}^\infty dxx^2f^\prime(x)=\frac{\pi^2}{3},
\end{equation}
we have
\begin{equation}
C_n=\frac{1}{2\pi^2}\left[(n+1)\left(\mu^2+\frac{\pi^2}{3}T^2\right)\rho^{2n}+\frac{1}{12}n(2n-1)\rho^{2n-2}\right] ,
\end{equation}
with $\rho=r\sin\theta$. Substituting into  Eq.~(\ref{eq:fa8}) and summing up the series, we end up with
\begin{equation}
 J_A^z =\left(\frac{1}{6}T^2+\frac{\mu^2}{2\pi^2}\right)\frac{\omega}{(1-\omega^2\rho^2)^2}+\frac{\omega^3}{24\pi^2}\frac{1+3\omega^2\rho^2}{(1-\omega^2\rho^2)^3},
\label{eq_ren5}
\end{equation}
which is in agreement with the closed-end formula derived in cylindrical coordinates in literature~\cite{Ambrus:2014uqa}. An alternative derivation in cylindrical coordinates is presented in Appendix B. To the cubic order in $\omega$,  Eq.~(\ref{eq_ren5}) yields the formula derived in Ref.~\cite{Ambrus:2014uqa}.
\begin{figure}[ht]
  \centering
\includegraphics[width=4.52in,height=3.09in]{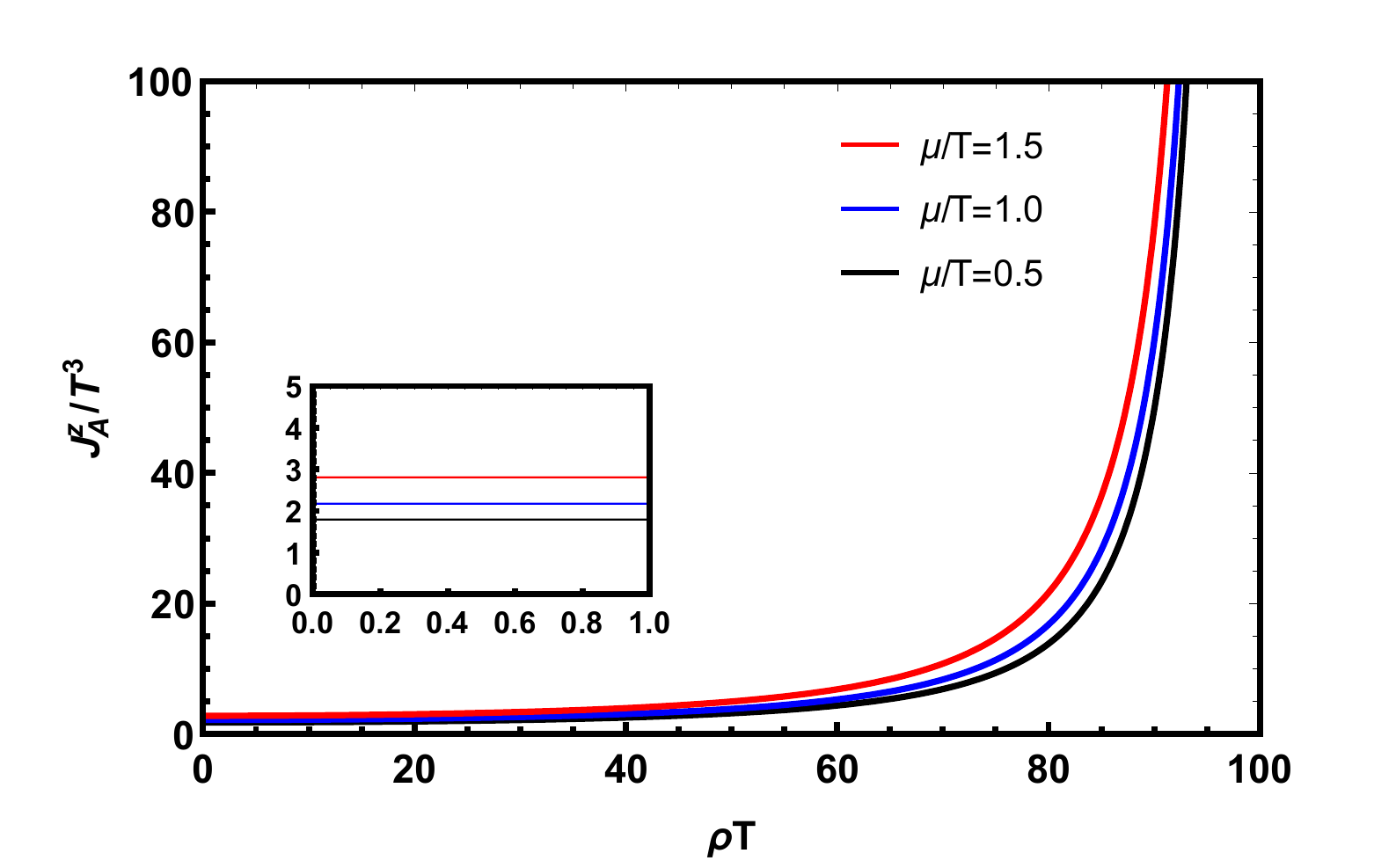}\\
  \caption{The ratio of axial vector current $J_A^z$ over $T^3$ of massless fermions far from the boundary in Eq.~(\ref{eq_ren5}) with the angular velocity $\omega=0.01\,T$ as a function of $\rho T$ where $\rho$ and T are radius and temperature, respectively. The black, blue, and red lines are with $\mu/T$ = 0.5, 1.0, and 1.5, respectively. The inner panel is for $\rho T$= 0--1.0.}              \label{eq:t7}
\end{figure}
As is shown in the step from  Eq.~(\ref{eq:c}) to  Eq.~(\ref{eq:c1}), the key reason for having the closed form of the axial current  Eq.~(\ref{eq_ren5}) is that the density of states for massless fermions is proportional to an integer power of the energy $E_k=k$ so that the integration by part terminates with a finite number of terms for arbitrary $n$. This is no longer the case for massive fermions.

 Eq.~(\ref{eq_ren5}) is plotted in Fig.~{\ref{eq:t7}}, where we set $\omega=0.01T$, a rough estimate of the vorticity of the QGP fireball created in RHIC. The pole at $\omega r\sin\theta=1$ occurs where the linear speed of rotation reaches the speed of light and the linear speed beyond the pole becomes superluminal which is not admissible. Therefore the Hamiltonian in  Eq.~(\ref{eq:pq1}) applies only to a finite volume, which in the case of the sphere discussed in this section requires its radius below $1/\omega$. The axial vector current in  Eq.~(\ref{eq_ren5}) is thereby free from the pole within the sphere but the finite size effect comes to play. Unless the finite size effect falls to zero faster than a power series in $r/R>\omega r$, its contribution will be of the same order of importance of the higher order terms of  Eq.~(\ref{eq_ren5}).

Coming to the transverse component far from the boundary, the typical contribution to the thermal average comes from $j<<kR$ and the sum over $k$ in Eq.~(\ref{eq:fa7}) and wave function normalization can be approximated by Eq.~(\ref{eq:sum_intgl}) and Eq.~(\ref{norm}). Following  Eq.~(\ref{j+2}), Eq.~(\ref{j+5}), Eq.~(\ref{j+6}) and Eq.~(\ref{j+7}) we end up with

\begin{eqnarray}
J_{A}^+ &=& \frac{2}{\pi}\int_0^\infty dkk^2\sum_{j,m}g_m(k)\left[j_{j-\frac{1}{2}}^2(kr)\eta_{j,j-\frac{1}{2},m}(\theta)+j_{j+\frac{1}{2}}^2(kr)\eta_{j,j+\frac{1}{2},m}(\theta)\right]\nonumber\\
&=& \frac{2}{\pi}\int_0^\infty dkk^2\left[\sum_{l=0}^\infty\frac{e^{-i\phi}}{2l+1}\sum_{m'}g_{m'+\frac{1}{2}}(k)\left[j_l^2(kr)Y_{lm'}^*(\theta,\phi)L_+Y_{lm'}(\theta,\phi)\right]  \right.\nonumber\\
& &\left. -\sum_{l=1}^\infty\frac{e^{-i\phi}}{2l+1}\sum_{m'}g_{m'+\frac{1}{2}}(k)\left[j_l^2(kr)Y_{lm'}^*(\theta,\phi)L_+Y_{lm'}(\theta,\phi)\right]\right] \nonumber\\
&=& \frac{2}{\pi}\int_0^\infty dkk^2e^{-i\phi}g_{\frac{1}{2}}(k)j_0^2(kr)Y_{00}^*(\theta,\phi)L_+Y_{00}(\theta,\phi)=0,
\label{longbulk}
\end{eqnarray}
where
\begin{equation}
g_m(k)\equiv\frac{1}{e^{\beta(k-m\omega-\mu)}+1}+\frac{1}{e^{\beta(k-m\omega+\mu)}+1}.
\label{gfunction}
\end{equation}
The absence of the transverse components is expected since the finite size effect can be neglected in the bulk and the spherical shape and cylindrical shape of the volume make no difference there.

\subsection{Axial vector current on the boundary}\label{sec:bm=0}
Moving to the boundary of a QGP fireball, we have to distinguish the radial momentum $k$ of the wave function $u_{kjlm}$ for $l=j-1/2$ and $l=j+1/2$ because of the different quantization conditions  Eq.~(\ref{MIT1}) and  Eq.~(\ref{MIT2}). It follows from Table~\ref{eq:t1} and  Eq.~(\ref{norm}) for $M=0$ that
\begin{equation}
u_{k,j,j\pm1/2,m}^\dagger\Sigma_3u_{k,j,j\pm1/2,m}=\frac{\Theta_{jm}(\theta)}{R^3\left(2\pm\frac{2j+1}{kR}\right)}.
\label{eq:boundary1}
\end{equation}
 The radial momentum $k$ of  Eq.~(\ref{eq:boundary1}) follows from  Eq.~(\ref{MIT1}) and  Eq.~(\ref{MIT2}). The axial vector current on the boundary is obtained upon substitution of  Eq.~(\ref{eq:boundary1}) into  Eq.~(\ref{eq:fa7}).
An analytic expression of the boundary axial vector current can be derived for the linear order term of the Taylor expansion in $\omega$ i.e. the chiral conductivity, at high temperature, i.e. $T\gg1/R$. We have
\begin{equation}
 J_A^z =-\frac{\omega}{R^3T}\left[\sum_{\lambda=\pm1,n,j}\frac{f^\prime\left(\frac{\lambda k_{nj}^{-}-\mu}{T}\right)}{2-\frac{2j+1}{k_{nj}^- R}}+\sum_{\lambda=\pm1,n,j}\frac{f^\prime\left(\frac{\lambda k_{nj}^{+}-\mu}{T}\right)}{2+\frac{2j+1}{k_{nj}^+ R}}\right]\sum_{m=-j}^j m\Theta_{jm}(\theta)+O(\omega^3),
\label{AVE1}
\end{equation}
where $k_{nj}^\mp$ stands for the solutions of  Eq.~(\ref{MIT1}) (``$-$'' sign) or   Eq.~(\ref{MIT2}) (``$+$'' sign). It follows from the definition  Eq.~(\ref{eq:Theta}) and the explicit form of the spinor spherical harmonics  Eq.~(\ref{eq:fa1}) that
\begin{equation}
\sum_{m=-j}^j m\Theta_{jm}(\theta)=\frac{2j+1}{4\pi}+\rho_{j-1/2}(\theta)-\rho_{j+1/2}(\theta),
\label{AVE2}
\end{equation}
 where
\begin{equation}\begin{aligned}
\rho_{l}(\theta)&\equiv\frac{2}{2l+1}\sum_{m^\prime=-l}^{l}(m^\prime)^{2}|Y_{lm^\prime}(\theta,\phi)|^{2} \\
&=\left.-\frac{2}{2l+1}\frac{d^{2}}{d\epsilon^{2}}\sum_{m^\prime=-l}^{l}Y_{lm^\prime}^{*}(\theta,\phi)Y_{lm^\prime}(\theta,\phi+\epsilon)\right|_{\epsilon=0} \\
&=\left.-\frac{1}{2\pi}\frac{d^{2}}{d\epsilon^{2}}P_{l}\left(1-2\sin^{2}\theta\sin^{2}\frac{\epsilon}{2}\right)\right|_{\epsilon=0}=\frac{l(l+1)}{4\pi}\sin^{2}\theta,
\label{AVE3}
\end{aligned}\end{equation}
and the addition formula of the spherical harmonics has been employed. Combining  Eq.~(\ref{AVE1}),  Eq.~(\ref{AVE2}) and  Eq.~(\ref{AVE3}), we arrive at
\begin{equation}
 J_{A}^{z}=-\frac{\omega}{4\pi R^{3}T}\cos^{2}\theta\sum_{\lambda=\pm1,n,j}\left[f^{\prime}\left(\frac{\lambda k_{nj}^{-}-\mu}{T}\right)\frac{2j+1}{2-\frac{2j+1}{k_{nj}^{-}R}}+f^{\prime}\left(\frac{\lambda k_{nj}^{+}-\mu}{T}\right)\frac{2j+1}{2-\frac{2j+1}{k_{nj}^{+}R}}\right]  .
\label{AVE4}
\end{equation}
    To evaluate the summation over $k$ and $j$ under the condition $T\gg 1/R$ or $\mu\gg 1/R$, we notice that $kR\gg 1$ and the wave functions of large $j$ become important because of the centrifugal force.  The asymptotic formula  Eq.~(\ref{eq:asymp}) is no longer sufficient to serve the purpose and one has to switch to the Debye formula~\cite{Wang2012} for Bessel function of large argument and large order \footnote{The Debye formula is instrumental to reproduce the same density of states in spherical coordinates as that derived in Cartesian coordinates under a periodic boundary condition~\cite{Lambert}.},
\begin{equation}
J_{\nu}(\nu\sec\beta)\cong\sqrt{\frac{2}{\nu\pi\tan\beta}}\cos\left(\nu\tan\beta-\nu\beta-\frac{\pi}{4}\right)\quad\ \ \nu\gg1 ,    \label{b1}
\end{equation}
which implies
\begin{equation}
j_{l}(kR)=\sqrt{\frac{\pi}{2kR}}J_{l+\frac{1}{2}}(kR)=\frac{1}{\left(l+\frac{1}{2}\right)\sqrt{\sec\beta \tan\beta}}\cos\left[\left(l+\frac{1}{2}\right)(\tan\beta-\beta)-\frac{\pi}{4}\right]  ,            \label{b2}
\end{equation}
for a spherical Bessel function. The MIT boundary conditions  Eq.~(\ref{MIT1}) and  Eq.~(\ref{MIT2}) becomes then:
\begin{equation}
\frac{1}{j\sqrt{\sec\beta \tan\beta}}\cos\left[j(\tan\beta-\beta)-\frac{\pi}{4}\right] = \pm \frac{1}{(j+1)\sqrt{\sec\beta^\prime \tan\beta^\prime}}\cos\left[(j+1)(\tan\beta^\prime-\beta^\prime)-\frac{\pi}{4}\right]  ,
\label{eq:debye}
\end{equation}
with
\begin{equation}
j\sec\beta=(j+1)\sec\beta^\prime=kR.
\label{eq:constraint}
\end{equation}
The large $j$ serves as the guideline to sort the order of approximation.
 Eq.~(\ref{eq:constraint}) gives rise to the leading order relation between $\beta$ and $\beta^\prime$
\begin{equation}
\beta^\prime=\beta-\frac{1}{j}\cot\beta.
\label{eq:prime}
\end{equation}
Substituting  Eq.~(\ref{eq:prime}) to RHS of  Eq.~(\ref{eq:debye}) and dropping the terms beyond the order of $1/j$, the boundary condition is reduced to
\begin{equation}
\cos\left[j(\tan\beta-\beta)-\frac{\pi}{4}\right]=\pm\cos\left[j(\tan\beta-\beta)-\frac{\pi}{4}-\beta\right],
\end{equation}
with the solutions
\begin{equation}
j(\tan\beta-\beta)-\frac{\beta}{2}=\left(n+\frac{1}{4}\right)\pi  ,
\end{equation}
for the upper sign and
\begin{equation}
j(\tan\beta-\beta)-\frac{\beta}{2}=\left(n+\frac{3}{4}\right)\pi ,
\end{equation}
for the lower sign, where $n$ is a positive integer. Together with relation between $\beta$ and the radial momentum $k$ in  Eq.~(\ref{eq:constraint}), we have~\cite{Lambert}
\begin{equation}
\delta n = \frac{R}{\pi}\sin\beta\delta k\simeq\frac{R}{\pi}\sqrt{1-\left(\frac{j}{kR}\right)^2}\delta k  ,
\end{equation}
to the leading order of large $j$ for both signs in  Eq.~(\ref{eq:debye}). Converting the summation over $k$ and $j$ in  Eq.~(\ref{AVE4}) to integrals, we end up with the leading order axial vector current on the boundary
\begin{equation}
\begin{aligned} J_{A}^{z}  & \cong-\frac{\omega}{4\pi^{2}R^{2}T}\cos^{2}\theta\int_{0}^{\infty}dk\sum_{j}\sqrt{1-\left(\frac{j}{kR}\right)^{2}}\left(\frac{j}{1-\frac{j}{kR}}+\frac{j}{1+\frac{j}{kR}}\right)\sum_{\lambda=\pm1}f^{\prime}\left(\frac{\lambda k-\mu}{T}\right)\\
 & \cong-\frac{\omega}{4\pi^{2}T}\cos^{2}\theta\int_{0}^{\infty}dkk^{2}\sum_{\lambda=\pm1}f^{\prime}\left(\frac{\lambda k-\mu}{T}\right)\int_{0}^{1}du\sqrt{1-u^{2}}\left(\frac{u}{1-u}+\frac{u}{1+u}\right)\\
 & =-\frac{\omega}{2\pi^{2}T}\cos^{2}\theta\int_{0}^{\infty}dkk^{2}\sum_{\lambda=\pm1}f^{\prime}\left(\frac{\lambda k-\mu}{T}\right)\int_{0}^{1}du\frac{u}{\sqrt{1-u^{2}}}\\
 & =-\frac{\omega}{2\pi^{2}T}\cos^{2}\theta\int_{0}^{\infty}dkk^{2}\sum_{\lambda=\pm1}f^{\prime}\left(\frac{\lambda k-\mu}{T}\right)=\left(\frac{1}{6}T^{2}\omega+\frac{\mu^2}{2\pi^2}\omega\right)\cos^{2}\theta  .
\end{aligned}     \label{b9}
\end{equation}
Therefore, the longitudinal axial vorticalconductivity vanishes along the equator, consistent with the general statement according to  Eq.~(\ref{eq:equator}) and matches the axial vorticalconductivity far from the boundary at the poles ($\theta=0,\pi$).

To the linear order in $\omega$, the transverse component of the axial vector current $\boldsymbol{J}_A$ is obtained by replacing $\Theta_{jm}(\theta)$ of the formula for the longitudinal component Eq.~(\ref{AVE1}) by
\begin{eqnarray}
\eta_{j,j-\frac{1}{2},m}(\theta)+\eta_{j,j+\frac{1}{2},m}(\theta)=\frac{1}{2j}e^{-i\phi}Y_{j-\frac{1}{2},m-\frac{1}{2}}^*(\theta,\phi)L_+Y_{j-\frac{1}{2},m-\frac{1}{2}}(\theta,\phi) \nonumber\\
-\frac{1}{2(j+1)}e^{-i\phi}Y_{j+\frac{1}{2},m-\frac{1}{2}}^*(\theta,\phi)L_+Y_{j+\frac{1}{2},m-\frac{1}{2}}(\theta,\phi),
\end{eqnarray}
i.e.
\begin{eqnarray}
J_A^+ &= -\frac{\omega}{R^3T}\left[\sum_{\lambda=\pm1,n,j}\frac{f^\prime\left(\frac{\lambda k_{nj}^{-}-\mu}{T}\right)}{2-\frac{2j+1}{k_{nj}^- R}}+\sum_{\lambda=\pm1,n,j}\frac{f^\prime\left(\frac{\lambda k_{nj}^{+}-\mu}{T}\right)}{2+\frac{2j+1}{k_{nj}^+ R}}\right]\nonumber\\
&\times \sum_{m=-j}^j m\left[\frac{1}{2j}e^{-i\phi}Y_{j-\frac{1}{2},m-\frac{1}{2}}^*(\theta,\phi)L_+Y_{j-\frac{1}{2},m-\frac{1}{2}}(\theta,\phi)  \right.\nonumber\\&\left.
-\frac{1}{2(j+1)}e^{-i\phi}Y_{j+\frac{1}{2},m-\frac{1}{2}}^*(\theta,\phi)L_+Y_{j+\frac{1}{2},m-\frac{1}{2}}(\theta,\phi)\right].\label{j1}
\label{AVEt1}
\end{eqnarray}
The summation over $m$ can be carried out similarly to Eq.~(\ref{AVE3}) and we find, with $m=m'+\frac{1}{2}$ and $l=j\pm\frac{1}{2}$, that
\begin{eqnarray}
&&\sum_{m'}\left(m'+\frac{1}{2}\right)Y_{lm'}^*(\theta,\phi)L_+Y_{lm'}(\theta,\phi) \nonumber\\
=&&i\lim_{(\theta',\phi')\to(\theta,\phi)}
e^{\frac{i}{2}\phi'}\frac{\partial}{\partial\phi'}e^{-\frac{i}{2}\phi'}L_+\sum_{m'}Y_{lm'}^*(\theta',\phi')Y_{lm'}(\theta,\phi)\nonumber\\
=&&i\frac{2l+1}{4\pi}\lim_{(\theta',\phi')\to(\theta,\phi)}e^{\frac{i}{2}\phi'}\frac{\partial}{\partial\phi'}e^{-\frac{i}{2}\phi'}\left(\frac{\partial}{\partial\theta}+i\cot\theta\frac{\partial}{\partial\phi}\right)\nonumber\\
&&.P_l(\cos\theta'\cos\theta+\sin\theta'\sin\theta\cos(\phi'-\phi))\nonumber\\
=&&-\frac{l(l+1)(2l+1)}{16\pi}\sin2\theta,\label{j2}
\end{eqnarray}
where the derivative formula of the Legendre polynomial
\begin{equation}
P_l^\prime(1)=\frac{l(l+1)}{2}
\end{equation}
is employed. Approximating the sum over $n$ snd $j$ by integrals of Eq.~(\ref{j1}) and Eq.~(\ref{j2}), we obtain the transverse component of the axial vector current
\begin{eqnarray}
 J_A^+ &=& \frac{\omega}{32\pi^2R^2}\sin2\theta\int_{0}^{\infty}dk\sum_{j}\sqrt{1-\left(\frac{j}{kR}\right)^{2}}\left(\frac{j}{1-\frac{j}{kR}}+\frac{j}{1+\frac{j}{kR}}\right)\sum_{\lambda=\pm1}f^{\prime}\left(\frac{\lambda k-\mu}{T}\right)\nonumber\\
&=&\frac{\omega}{32\pi^2}\sin2\theta\int_{0}^{\infty}dkk^{2}\sum_{\lambda=\pm1}f^{\prime}\left(\frac{\lambda k-\mu}{T}\right)\int_{0}^{1}du\sqrt{1-u^{2}}\left(\frac{u}{1-u}+\frac{u}{1+u}\right) \nonumber\\
&=&\frac{1}{16\pi^2}\left(\frac{\pi^2}{3}T^2+\mu^2\right)\omega\sin2\theta
\end{eqnarray}
which is of the same order of magnitude as the longitudinal component. Restoring the cylindrical coordinates via
\begin{equation}
\cos\theta=\frac{z}{\sqrt{\rho^2+z^2}} \qquad \sin\theta=\frac{\rho}{\sqrt{\rho^2+z^2}}
\end{equation}
we have
\begin{equation}
J_A^+=\frac{1}{8\pi^2}\left(\frac{\pi^2}{3}T^2+\mu^2\right)\frac{\rho z}{\sqrt{\rho^2+z^2}}
\end{equation}
which is independent of the azimuthal angle and odd in $z$, consistent with the symmetry argument in Sec.~\ref{sec:12}.

\section{axial chiral vortical effect of massive fermions with finite-size effect}\label{sec:4}
\subsection{Mass correction of axial vector current far from the boundary}\label{sec:m}
For massive fermions, the same approximation of the MIT boundary condition applied to massless fermions reduces the axial vector current $\boldsymbol{J}_{A}$ in Eq.~(\ref{eq:fa7}) far from the boundary to
\begin{equation}
\mathcal{J}_A=\frac{R}{2 \pi} \int_{-\infty}^{\infty} d k \sum_{j l m}\left[\frac{1}{e^{\beta(E_{k}-m \omega-\mu)}+1}-\frac{1}{e^{\beta(E_{k}+m \omega-\mu)}+1}\right] u_{k j l m}^{\dagger} \boldsymbol{\Sigma} u_{k j l m}  ,
\label{y1}
\end{equation}
with $E_k=\sqrt{k^2+M^2}$. As $dkk^2=dE_kE_k\sqrt{E_k^2-M^2}$, the density of states is no longer an integer power of the energy $E_k$ and a closed-end formula like Eq.~(\ref{eq_ren5}) does not exist. We shall stay with the linear response of $J_A^z$ to $\omega$ in what follows and calculate the axial vertical conductivity. It is straightforward to verify that the combination
\begin{equation}
u_{k, j, j-1/2, m}^{\dagger} \boldsymbol{\Sigma} u_{k, j, j-1/2, m}+u_{k, j, j+1/2, m}^{\dagger} \boldsymbol{\Sigma} u_{k, j, j+1/2, m}  ,
\end{equation}
with the radial wave functions in Table~\ref{eq:t1} and the normalization constant Eq.~(\ref{norm}) at a given $k$ is independent of the mass $M$ and thereby takes the same massless form. For the longitudinal component, the spinor spherical harmonics part can be reduced the same way as in Sec.~\ref{sec:m=0} and Eq.~(\ref{y1}) becomes, to the order $\omega$,
\begin{equation}
\begin{aligned}
J_{A}^z=&-\frac{4\omega}{\pi T}\int_0^{\infty}dkk^{2}\left[f^{'}\left(\frac{E-\mu}{T}\right)+f^{'}\left(\frac{E+\mu}{T}\right)\right]\sum_{l=0}^{\infty}j_{l}^{2}(kr)\sum_{m^{\prime}=-l}^{l}\left(m^{\prime}+\frac{1}{2}\right)\left|Y_{lm^{\prime}}(\theta,\varphi)\right|^{2}  . \\
	\end{aligned}	  \label{y2}
\end{equation}
Using the relation of Eq.~(\ref{s1}), Eq.~(\ref{y2}) becomes
\begin{equation}
J_{A}^z = \frac{1}{2\pi^{2}}T^{2}\omega\int_{M/T}^{\infty}\sum_{\lambda=\pm1}\lambda x\sqrt{(\lambda x)^{2}-(\frac{M}{T})^{2}}\frac{e^{\lambda x-\mu/T}}{(e^{\lambda x-\mu/T}+1)^{2}}dx,
\label{eq:qd15}
\end{equation}
where we have transformed the integration variable from $k$ to $x=E/T$ with $E=\sqrt{k^2+M^2}$. The integral Eq.~(\ref{eq:qd15}) can be converted to a contour integral by the observation that
\begin{equation}
\begin{aligned}
&\int_{M/T}^{\infty}\sum_{\lambda=\pm1}\lambda x\sqrt{(\lambda x)^{2}-(\frac{M}{T})^{2}}\frac{e^{\lambda x-\mu/T}}{(e^{\lambda x-\mu/T}+1)^{2}}dx \\
=&\mathrm{Re}\left[\int_{-\infty+i0^{+}}^{\infty+i0^{+}}z\sqrt{z^{2}-(\frac{M}{T})^{2}}\frac{e^{z-\mu/T}}{(e^{z-\mu/T}+1)^{2}}dz\right]  \\
=&\mathrm{Re}[I+I'] ,
\end{aligned}	     \label{eq:qd16}
\end{equation}
where the first two terms of the Taylor expansion of $\sqrt{z^2-\frac{M^2}{T^2}}$ in the powers of $M$ is included in $I^\prime$, i.e.,
\begin{equation}
\begin{aligned}
\mathrm{Re}[I']	=&\mathrm{Re}\left[\int_{-\infty+i0^{+}}^{\infty+i0^{+}}[z^{2}-\frac{a^{2}}{2}]\frac{e^{z-\mu/T}}{(e^{z-\mu/T}+1)^{2}}dz\right]   \\
=&\left[\int_{-\infty}^{\infty}[(x+\frac{\mu}{T})^{2}-\frac{a^{2}}{2}]\frac{e^{x}}{(e^{x}+1)^{2}}dx\right]   \\
		=&\frac{\pi^{2}}{3}+\frac{\mu^2}{T^2}-\frac{a^{2}}{2}  ,
\end{aligned}	    \label{eq:qd17}
\end{equation}
with $a=\frac{M}{T}$. Then the integrand of $I$ vanishes sufficiently fast at infinity so that the integration path can be closed from infinity on the upper or lower $z$-plane and the integral equals to the sum of residues at the poles of the distribution function within the contour. Closing the path from the upper plane, we have the poles
\begin{equation}
z=\frac{\mu}{T}+(2n+1)i\pi\equiv iv_n  ,
\end{equation}
within the contour, i.e., $n=0,1,2,...$. Consequently
\begin{equation}
\begin{aligned}
I=&\int_{-\infty+i0^{+}}^{\infty+i0^{+}}[z\sqrt{z^{2}-a^{2}}-z^{2}+\frac{a^{2}}{2}]\frac{e^{z-\mu/T}}{(e^{z-\mu/T}+1)^{2}}dz , \\
&=2\mathrm{Re}\left[\pi\sum_{n=0}^\infty v_n\left((1+\frac{a^{2}}{v_n^{2}})^{\frac{1}{2}}+(1+\frac{a^{2}}{v_n^{2}})^{-\frac{1}{2}}-2\right)\right],
\end{aligned}	    \label{eq:qd18}
\end{equation}
Combining with Eq.~(\ref{eq:qd15}), Eq.~(\ref{eq:qd16}) and Eq.~(\ref{eq:qd17}), we have
\begin{equation}
J_{A}^z=\sigma\omega,
\end{equation}
with the axial vertical conductivity of massive fermions
\begin{equation}
\sigma=\frac{1}{6}T^{2}+\frac{\mu^2}{2\pi^2}+\frac{T^2}{\pi^2}\mathrm{Re}\left[\pi\sum_{n=0}^\infty v_n\left((1+\frac{a^{2}}{v_n^{2}})^{\frac{1}{2}}+(1+\frac{a^{2}}{v_n^{2}})^{-\frac{1}{2}}-2\right)\right].
\label{eq:sigma}
\end{equation}
Binomial expansions of the square roots in Eq.~(\ref{eq:sigma}), enable us to write
\begin{equation}
\sigma=\frac{1}{6}T^{2}+\frac{\mu^2}{2\pi^2}-\frac{M^{2}}{4\pi^{2}}+T^{2}\sum_{r=2}^{\infty}\frac{[(r-1)(2r-3)!!](-1)^{r}}{r!2^{r-1}}(2\pi)^{-2r}\zeta\left(2r-1,\frac{1}{2}+\frac{b}{2\pi i}\right)a^{2r},
\label{eq:qd20}
\end{equation}
where $\zeta(...)$ denotes the Hurwitz zeta function, defined by
\begin{equation}
\zeta(s,b)=\sum_{n=0}^\infty\frac{1}{(n+b)^s}  .
\end{equation}
Away from the branch points of the square roots in the summands, the infinite series Eq.~(\ref{eq:qd18}) converges uniformly with respect to $a$ and thereby the radius of convergence of the power series Eq.~(\ref{eq:qd20}) corresponds to the absolute value of the closest branch point to the origin of the complex $a$-plane, i.e. $\sqrt{\pi^2+(\mu/T)^2}$. This can also be inferred from the asymptotic behavior of the expansion coefficients of Eq.~(\ref{eq:qd20}). We also get Eq.~(\ref{eq:qd20}) in Appendix~\ref{sec:cylindrical} by using a cylindrical coordinate system and in Appendix~\ref{Dimensional} by Kubo formula via thermal diagram, which shows that this result, derived by different methods is robust. In particular, the thermal diagram requires UV regularization but the result is independent of regularization schemes.

At zero temperature, the summation over $n$ in Eq.~(\ref{eq:sigma}) can be converted to an integral and we obtain that
\begin{equation}\begin{array}{c}
\sigma=\frac{\mu^{2}}{2\pi^{2}}-\frac{M^{2}}{4\pi^{2}}+\int_{-i\mu}^{\infty-i\mu}d\xi\left(\sqrt{\xi^{2}+M^{2}}+\frac{\xi^{2}}{\sqrt{\xi^{2}+M^{2}}}-2\xi\right)=\begin{cases}
0 & (\mu<M)\\
\frac{1}{2\pi^{2}}\mu\sqrt{\mu^{2}-M^{2}} & (\mu>M)
\end{cases}.
\end{array}
\label{massive_bulk}\end{equation}
The zero $\sigma$ for $\mu<M$ is obvious from Eq.~(\ref{y2}) where the derivative of the distribution function vanishes exponentially in the limit $T\to 0$ for all $k$. The case with $\mu>M$ returns the massless result derived in Sec.~\ref{sec:m=0} for $M=0$.

\begin{figure}[ht]
  \centering
\includegraphics[width=4.52in,height=3.09in]{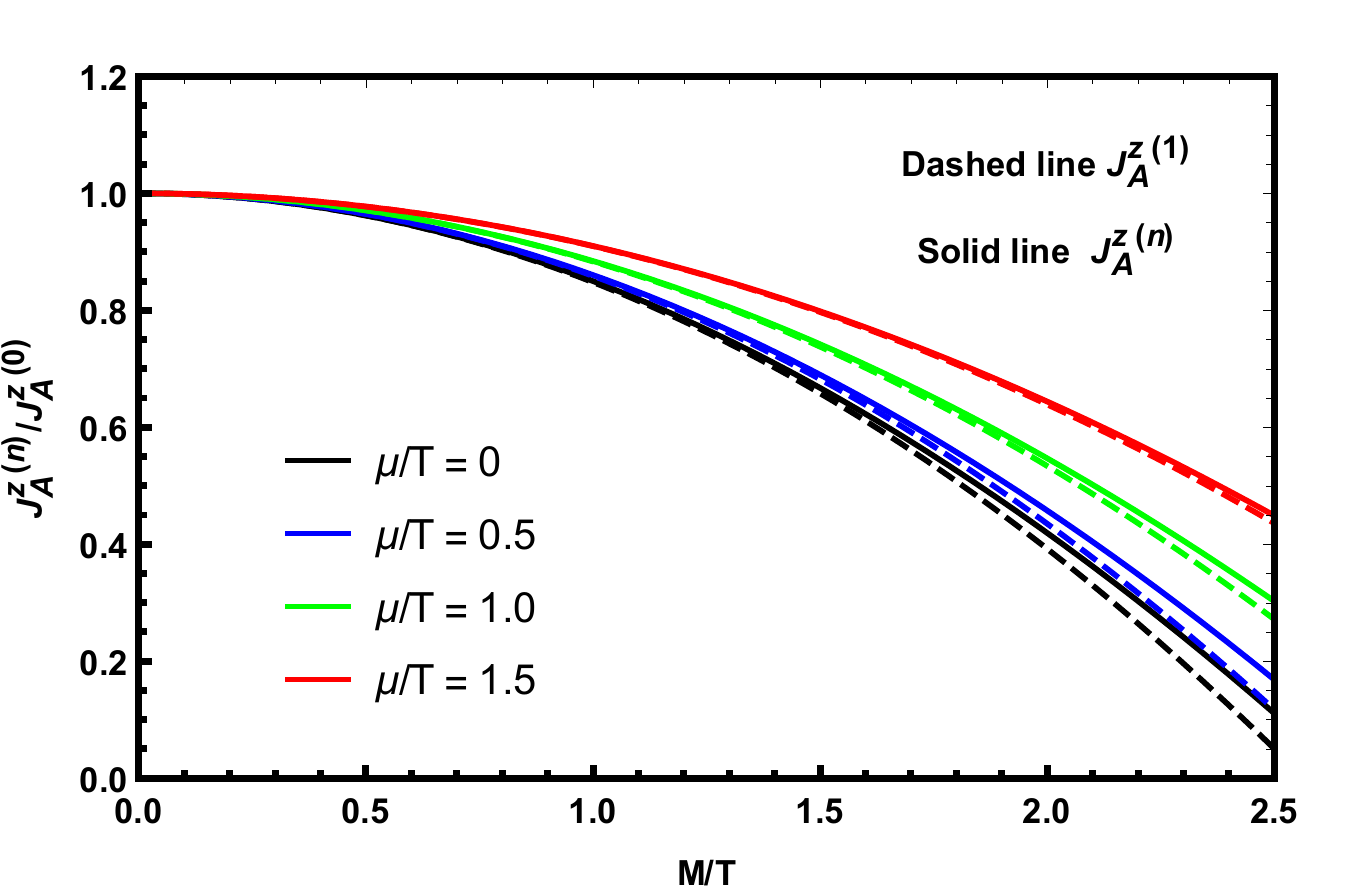}\\
  \caption{The ratio of axial vector current $J_{A}^{z(n)}$ including mass correction up to $M^{2n}$ over massless current $J_{A}^{z(0)}$ ($J_{A}^{z(n)}/J_{A}^{z(0)}$) as a function of the product of $M$ over $T$ ($M/T$), where black, blue, green, and red lines are with $\mu/T$ =0, 0.5, 1.0, and 1.5, respectively. The dashed lines are $J_{A}^{z(1)}/J_{A}^{z(0)}$ and the solid lines are $J_{A}^{z(n)}/J_{A}^{z(0)}$.}             \label{eq:t5}
\end{figure}
The axial vector current with mass correction is plotted in Fig.~\ref{eq:t5}, where the solid line is $J_{A}^{z(n)}$ and the dashed line is $J_{A}^{z(1)}$.  Here, $J_{A}^{z(0)}$ is the axial vector current at $M=0$, $J_{A}^{z(1)}$ is the axial vector current with only $M^{2}$ correction (The result of $J_{A}^{z(1)}$ is also obtained in Ref.~\cite{Lin:2018aon}), and $J_{A}^{z(n)}$ is the result including mass correction up to $M^{2n}$.  Their concrete expressions are
\begin{equation}
\begin{aligned}
J_{A}^{z(0)}&=\left(\frac{T^{2}}{6}+\frac{\mu^2}{2\pi^2}\right)\omega, \qquad   J_{A}^{z(1)}= \left(\frac{T^{2}}{6}+\frac{\mu^2}{2\pi^2}\right)\omega-\frac{M^{2}}{4\pi^{2}}\omega            , \\
J_{A}^{z(n)}&=\left(\frac{T^{2}}{6}+\frac{\mu^2}{2\pi^2}\right)\omega-\frac{M^{2}}{4\pi^{2}}\omega+T^{2}\omega\sum_{r=2}^{n}A_{r}M^{2r}                   . \label{eq:t4}
\end{aligned}\end{equation}

We can see clearly that $J_{A}^{z(n)}/J_{A}^{z(0)}$ decreases with $M/T$. This is because the presence of mass generally inhibits the fluidity, thus suppressing the vortical conductivity. While the presence of chemical potential slows down this inhibition, when we fix $M/T$, $J_{A}^{z(n)}/J_{A}^{z(0)}$ and $J_{A}^{z(1)}/J_{A}^{z(0)}$ are all increase with increasing $\mu/T$.

Take $s$ quark as an example. We set $M=150$MeV, $\mu/T=1.0$, $n=2000$, and show the numerical values of the mass correction in Table~\ref{eq:t3}.

\begin{table}[hpt]
\caption{Mass correction of axial current when $M=150$MeV, $\mu/T=1.0$, $n=2000$. }
\centering
\begin{tabular}{|c|c|c|c|}
\hline $T(\mathrm{MeV})$ & $J_{A}^{z(1)} / J_{A}^{z(0)}$ & $J_{A}^{z(n)} / J_{A}^{z(0)}$ & $\left(T^2\omega\sum_{r=2}^{n}A_{r}M^{2r} \right) /J_{A}^{z(0)}$ \\
\hline $100 $ & $ 0.737754 $ & $ 0.741961 $ & $ 4.20713\times10^{-3} $ \\
\hline $150 $ & $ 0.883446 $ & $ 0.884285 $ & $ 8.38337\times10^{-4} $ \\
\hline $200 $ & $ 0.934439 $ & $ 0.934704 $ & $ 2.65816\times10^{-4} $ \\
\hline $250 $ & $ 0.958041 $ & $ 0.95815  $ & $ 1.08964\times10^{-4} $ \\
\hline
\end{tabular}       \label{eq:t3}
\end{table}
Far from the boundary, the mass correction for $s$ quark is modest for the selected temperature and chemical potential and is dominated by the leading order $O(M^2)$ correction. On the boundary, the leading order mass correction is $O(M)$ as shown below, The mass suppression for $s$ quark is thereby much stronger there.

For the transverse component of the axial vector current of massive fermions far from the boundary, all we need is to replace $k$ in $g_m(k)$ of (\ref{longbulk}) with $E_k$ and the result remains zero, the same as the massless case.

\subsection{Mass correction of axial vector current on the boundary}\label{sec:bm}
An analytical result can also be obtained for the leading order mass correction on the spherical boundary under the same approximation of Sec.~\ref{sec:bm=0}, i.e. $T\gg 1/R$. For massive fermions, it follows from Eq.~(\ref{eq:norm}) that Eq.~(\ref{eq:boundary1}) is replaced by
\begin{equation}
u_{k,j,j\pm1/2,m}^\dagger\Sigma_3u_{k,j,j\pm1/2,m}=\frac{\Theta_{jm}(\theta)}{2R^3b\left(b\pm\frac{j}{kR}\right)},
\label{eq:boundary1m}
\end{equation}
with $b=E_k/\sqrt{E_{k}^2-M^2}$, where we have substituted Eq.~(\ref{tangent}) for the trigonometric functions in the normalization constant Eq.~(\ref{eq:norm}) and made the approximation $2j+2\simeq 2j$ in the last term inside the parentheses for large $j$. The conversion from the sum of the radial momentum into an integral proceeds the same way as for the massless case in Sec.~\ref{sec:bm=0} and we end up with the following form of the axial vector current to the order $O(\omega)$
\begin{equation}
J_{A}^{z}=-\frac{\omega}{4\pi^{2}T}\cos^{2}\theta\int_{0}^{\infty}dkk^{2}\sum_{\lambda=\pm1}f^{\prime}\left(\frac{\lambda E_k-\mu}{T}\right)\frac{1}{b}\int_{0}^{1}du\sqrt{1-u^{2}}\left(\frac{u}{b-u}+\frac{u}{b+u}\right).
\label{axial_mass}
\end{equation}
The integration over $u$ can be carried out readily
\begin{eqnarray}
\frac{1}{b}\int_0^1 du\sqrt{1-u^{2}} \left(\frac{u}{b-u}+\frac{u}{b+u}\right) &=&2-2\sqrt{b^2-1}\tan^{-1}\frac{1}{\sqrt{b^2-1}}\nonumber\\
&=& 2-\frac{\pi M}{E_k}+O\left(\frac{M^2}{E_k^2}\right).
\label{integral_b}
\end{eqnarray}
Consequently, the leading order mass correction is $O(M)$, stronger than $O(M^2)$ for the mass correction far from the boundary. Substituting Eq.~(\ref{integral_b}) into Eq.~(\ref{axial_mass}) and setting $E_k=k$ we find
\begin{equation}
J_{A}^{z}=J_{A}^{z(0)}+J_{A}^{z(1)}+...  ,\label{eq:B}
\end{equation}
where the first term, $J_{A}^{z(0)}$, is the axial-vector current of massless fermions, given by Eq.~(\ref{b9})
and the leading order mass correction reads
\begin{eqnarray}
J_{A}^{z(1)} &=& \frac{M\omega}{4\pi T}\cos^2\theta\int_0^\infty dkk\left[f^\prime\left(\frac{k-\mu}{T}\right)+f^\prime\left(\frac{k+\mu}{T}\right)\right]\nonumber \\
&=& -\frac{M\omega}{4\pi}\left[\mu+2T\ln \left(1+e^{-\frac{\mu}{T}}\right)\right]\cos^2\theta,  \label{eq:B1}
\end{eqnarray}
which is an even function of $\mu$. Adding Eqs.~(\ref{b9},~\ref{eq:B1}), we have the longitudinal axial vector current on the boundary up to the leading order mass correction.
\begin{equation}
\begin{aligned}
J_{A}^{z(B)}&=\left\{\frac{T^{2}}{6}+\frac{\mu^2}{2\pi^2}-\frac{M}{4\pi}\left[\mu+2T\ln \left(1+e^{-\frac{\mu}{T}}\right)\right]\right\}\omega\cos^2\theta                , \label{eq:t4}
\end{aligned}\end{equation}
where $J_{A}^{z(B)}$ is the axial vector current with only leading order mass correction on the boundary. We can see clearly that the mass correction is stronger on the boundary than that far from the boundary. The coefficient of $\omega$ of Eq.~(\ref{eq:t4}) gives rise to the axial vorticalconductivity on the boundary, Eq.~(\ref{anisotropicity1}) announced in the introduction. As $T\to 0$,
\begin{equation}
\frac{1}{T}f^\prime\left(\frac{\lambda E_k-\mu}{T}\right)\to \delta\left(\lambda E_k-\mu\right).
\end{equation}
With the aid of the integral Eq.~(\ref{integral_b}) together with the definition of $b$, we end up with a closed-end formula of the axial vorticalconductivivity to all orders of mass on the boundary
\begin{equation}\begin{array}{c}
\sigma=\begin{cases}
0 & (\mu<M)\\
\frac{1}{2\pi^{2}}\mu\sqrt{\mu^{2}-M^{2}}\left(1-\frac{M}{\sqrt{\mu^2-M^2}}\tan^{-1}\frac{\sqrt{\mu^2-M^2}}{M}\right)\cos^2\theta & (\mu>M)
\end{cases}.
\end{array}
\label{massive_boundary}\end{equation}
in parallel to Eq.~(\ref{massive_bulk}) in the bulk.

It is straightforward to extend the above analysis to the transverse component. Starting with  Eq.~(\ref{j+4}) and going through the gymnastics from  Eq.~(\ref{axial_mass}) to  Eq.~(\ref{eq:t4}) with $\cos^2\theta$ replaced by $\frac{1}{8}\sin2\theta$, we find the transverse axial vector current on the boundary up to the leading order of mass correction, i.e.
\begin{equation}
\begin{aligned}
J_{A}^{+(B)}&=\left\{\frac{T^{2}}{48}+\frac{\mu^2}{16\pi^2}-\frac{M}{24\pi}\left[\mu+2T\ln \left(1+e^{-\frac{\mu}{T}}\right)\right]\right\}\omega\sin2\theta.                 \label{eq:t4t}
\end{aligned}\end{equation}
At zero temperature, we have
\begin{equation}\begin{array}{c}
J_{A}^{+(B)}=\begin{cases}
0 & (\mu<M)\\
\frac{1}{16\pi^{2}}\mu\sqrt{\mu^{2}-M^{2}}\left(1-\frac{M}{\sqrt{\mu^2-M^2}}\tan^{-1}\frac{\sqrt{\mu^2-M^2}}{M}\right)\omega\sin2\theta & (\mu>M)
\end{cases}.
\end{array}
\label{massive_boundaryt}\end{equation}
valid to all orders in the mass $M$.

\section{Concluding Remarks}\label{sec:5}
Let us recapitulate what we accomplished in this work. We start with a general discussion of the axial vortical effect from symmetry perspectives and investigated the axial vortical effect of a free Dirac field in a finite sphere rotating with a given angular velocity $\omega$. For massless fermions far from the boundary, we were able to reproduce the closed-end formula derived within a cylinder in literature. On the boundary, the axial vector current displays both longitudinal and transverse components with respect to the rotation axis and the magnitude of each component depends on the colatitude angle of the spherical coordinates. For massive fermions, we get the mass correction of the chiral conductivity far from and on the boundary. In the former case, we expanded the chiral conductivity to all orders of mass with the leading order correction in agreement with what was reported in the literature. In the latter case, we found that the leading order mass correction is stronger than that of the former, $O(M)$ versus $O(M^2)$. To our knowledge, the axial vortical effect on the boundary, especially the emergence of the transverse component, has not been explored in literature.

\begin{figure}[ht]
	\centering
	\includegraphics[width=4.52in,height=3.09in]{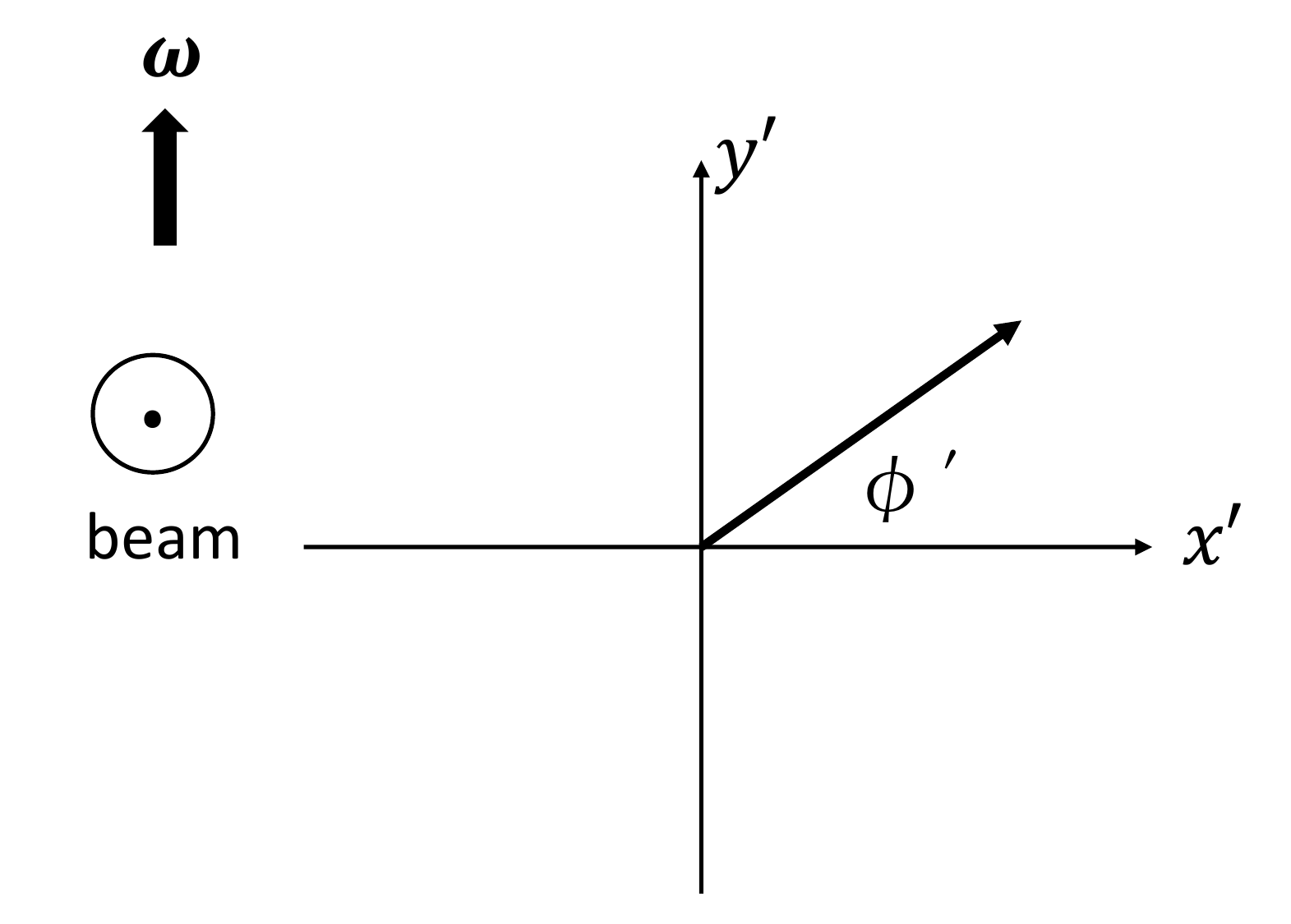}\\
	\caption{$x'$ and the beamline define the reaction plane.}              \label{beamline}
\end{figure}

While the value of the above results are mainly theoretical and cannot describe quantitatively the ACVE of a strongly interacting and expanding fireball of QGP, some qualitative speculations on the finite size effect in heavy ion collisions remain instructive. The quadrupole factor $\cos^2\theta$ in Eq.~(\ref{anisotropicity1}) would suppress the global polarization (z-component of Eq.~(\ref{AVEvector})) and
the perpendicular component in Eq.~(\ref{AVEvector}) would contribute to the polarization in the reaction plane shown in Fig.~\ref{beamline}, e.g. the longitudinal polarization (the polarization along the beam).

 To see the latter effect clearly, we assume that the beam is along  $\hat{\boldsymbol{y}}$, and rotate the coordinate system by $90^\circ$ around the x-axis, i.e. $y=-z^\prime=-r\cos\theta^\prime$, $z=y^\prime=r\sin\theta^\prime\sin\phi^\prime$ and $x=x^\prime=r\sin\theta^\prime\cos\phi^\prime$ with $r$ the radial coordinate. In terms of the polar angle, $\theta^\prime$ and azimuthal angle $\phi^\prime$ associated with the primed coordinates, the longitudinal component in Eq.~(\ref{AVEvector}) and Eq.~(\ref{anisotropicity1}) takes the form
\begin{equation}
\boldsymbol{J}_A\cdot\boldsymbol{\hat{z}}^\prime=-b\sin2\theta^\prime\sin\phi^\prime
\label{map}
\end{equation}
with $b=\left\{\frac{\mu^2}{16\pi^2} + \frac{1}{48}T^2-\frac{M}{32\pi}\left[\mu+2T\ln \left(1+e^{-\frac{\mu}{T}}\right)\right]\right\}\omega$. As the fragment hadrons, e. g. $\Lambda$ hyperons originated from the boundary layer are more likely flying in the radial direction, Eq.~(\ref{map}) maps out the longitudinal polarization profile of these hadrons with $\phi^\prime$ the angle of the transverse momentum with respect to the reaction plan, and $\theta^\prime$ related to the pseudorapidity via $\eta=-\ln\tan\frac{\theta^\prime}{2}$.

More investigations are required for the finite size effect discovered in this work to be practical with respect to the phenomenology of heavy ion collisions. These include exploring ACVE with the solution of the Dirac equation in an expanding sphere and/or incorporating the anisotropic ACVE conductivity in Eq.~(\ref{AVEvector}) into hydrodynamic models. We hope to report the progress along this line in the near future.

\section*{Acknowledgments}
We thank Ren-Da Dong and Xin-Li Sheng for fruitful discussions. This work is supported by the NSFC Grant Nos. 11735007, 11890710, 11890711, and 11890713.
\appendix
\section{ Axial vector current along the equator}\label{appendix:a}
To prove Eq.~(\ref{eq:equator}), we substitute the explicit form of $Z_{j,l,m}^{\dagger}(\theta,\phi)$ into Eq.~(\ref{eq:Theta}), i.e.
\begin{equation}
\begin{aligned}
\Theta_{jm}(\theta) =& \frac{1}{2j}\bigg{[}(j+m)\left|Y_{j-\frac{1}{2},m-\frac{1}{2}}(\theta,\phi)\right|^{2}-(j-m)\left|Y_{j-\frac{1}{2},m+\frac{1}{2}}(\theta,\phi)\right|^2\bigg{]}   \\
+&\frac{1}{2(j+1)}\bigg{[}(j-m+1)\left|Y_{j+\frac{1}{2},m-\frac{1}{2}}(\theta,\phi)\right|^{2}\\
-&(j+m+1)\left|Y_{j+\frac{1}{2},m+\frac{1}{2}}(\theta,\phi)\right|^2\bigg{]}  .
\end{aligned}
\end{equation}
As $\Theta_{jm}(\theta)$ is odd in $m$, we only need to consider the case with $m>0$. Setting $j=l+1/2$ and $m=\mu+1/2$, and using the expression of spherical harmonics in terms of the associated Legendre function, we have
\begin{equation}
\begin{aligned}
\Theta_{jm}(\theta)=&\frac{1}{4\pi(l+\mu+1)}\frac{(l-\mu)!}{(l+\mu)!}[(l+\mu+1)^{2}P_{l}^{\mu}(\cos\theta)^{2} \\
&-P_{l}^{\mu+1}(\cos\theta)^{2}+(l-\mu+1)^{2}P_{l+1}^{\mu}(\cos\theta)^{2}-P_{l+1}^{\mu+1}(\cos\theta)^{2}]    ,
\end{aligned}               \label{b5}
\end{equation}
with $\mu\ge 0$. It follows from the generating function of Legendre polynomials
\begin{equation}
\frac{1}{\sqrt{1-2zt+t^2}}=\sum_{l=0}^\infty t^lP_l(z),
\end{equation}
and the definition
\begin{equation}
P_l^\mu(z)=(-)^\mu(1-z^2)^{\frac{\mu}{2}}\frac{d^\mu P_l(z)}{dz^\mu},
\end{equation}
that
\begin{equation}
(-)^\mu(2\mu-1)!!(1-z^2)^{\frac{\mu}{2}}t^\mu(1-2zt+t^2)^{-\frac{1}{2}-\mu}=\sum_{l=\mu}^\infty t^lP_l^\mu(z).
\end{equation}
Setting $z=0$ and comparing the coefficients of $t^l$ on both sides, we obtain that~\cite{Gradshteyn2015}
\begin{equation}
P_{l}^{\mu}(0)=\frac{2^{\mu} \sqrt{\pi}}{\Gamma\left(\frac{l-\mu}{2}+1\right) \Gamma\left(\frac{-l-\mu+1}{2}\right)}.
\end{equation}
It is straightforward to verify that
\begin{equation}
P_{l+1}^{\mu+1}(0)=-(l+\mu+1)P_l^\mu(0),
\end{equation}
and
\begin{equation}
P_l^{\mu+1}(0)=(l-\mu+1)P_{l+1}^\mu(0).
\end{equation}
Eq.~(\ref{eq:equator}) is thereby proved.
\section{ Axial vector current in cylindrical coordinate system}
\label{sec:cylindrical}In this appendix, we first solve the free
Dirac equation in a cylindrical coordinate system, and then calculate
the axial vector current of the system of massive Dirac fermions which
uniformly rotates with angular velocity $\boldsymbol{\omega}=\omega\boldsymbol{e}_{z}$
along $z$-axis. We consider only the axial vector current far from the boundary and thereby ignore the finite size effect.
\subsection{Solution of the free Dirac equation in cylindrical coordinate system}
\label{subsec:free}We work in the chiral representation of gamma
matrices as adopted in Ref. \citep{Peskin:1995},
\begin{equation}
\gamma^{0}=\left(\begin{array}{cc}
0 & 1\\
1 & 0
\end{array}\right),\ \ \gamma^{i}=\left(\begin{array}{cc}
0 & \sigma_{i}\\
-\sigma_{i} & 0
\end{array}\right),\ \ \gamma^{5}=\left(\begin{array}{cc}
-1 & 0\\
0 & 1
\end{array}\right),\label{eq:p0}
\end{equation}
with $\sigma_{i}$ $(i=x,y,z)$ the three Pauli matrices.
The equation of motion for the free Dirac field $\Psi(t,\boldsymbol{r})$
can be written as
\begin{equation}
i\frac{\partial}{\partial t}\Psi(t,\boldsymbol{r})=\hat{H}\Psi(t,\boldsymbol{r}),\label{eq:qq0}
\end{equation}
with the Hamiltonian $\hat{H}=-i\gamma^{0}\boldsymbol{\gamma}\cdot\nabla+\gamma^{0}M$,
and the Dirac fermion mass $M$. Suppose that $\Psi(t,\boldsymbol{r})$
is an energy eigenstate with eigenvalue $E$, i.e. $\Psi(t,\boldsymbol{r})=e^{-iEt}\psi(\boldsymbol{r})$,
then Eq.~(\ref{eq:qq0}) becomes
\begin{equation}
\hat{H}\psi(\boldsymbol{r})=E\psi(\boldsymbol{r}),\label{eq:qq1}
\end{equation}
which is the energy eigenvalue equation of the Hamiltonian. It can
be proved that, these four Hermitian operators, $\hat{H},\hat{p}_{z},\hat{J}_{z},\boldsymbol{\Sigma}\cdot\hat{\boldsymbol{p}}$,
are commutative with each other, where $\boldsymbol{\Sigma}=\mathrm{diag}\,(\boldsymbol{\sigma},\boldsymbol{\sigma})$,
$\hat{\boldsymbol{p}}=-i\nabla$, $\hat{\boldsymbol{J}}=\boldsymbol{r}\times\hat{\boldsymbol{p}}+\frac{1}{2}\boldsymbol{\Sigma}$,
and $\hat{p}_{z},\hat{J}_{z}$ are the $z$-components of $\hat{\boldsymbol{p}}$
and $\hat{\boldsymbol{J}}$ respectively. In the following, we will
calculate the common eigenstates of these four operators in cylindrical
coordinate system.
We set $\psi=(\psi_{1},\psi_{2})^{T}$, where $\psi_{1},\psi_{2}$
are both two-component spinors, then Eq.~(\ref{eq:qq1}) can be replaced
by following two equations,
\begin{equation}
(\nabla^{2}+E^{2}-M^{2})\psi_{1}=0,\label{eq:qq2}
\end{equation}
\begin{equation}
\psi_{2}=\frac{1}{M}\left(E-i\boldsymbol{\sigma}\cdot\nabla\right)\psi_{1}.\label{eq:p1}
\end{equation}
In a cylindrical coordinate system, the form of $\nabla^{2}$ is
\begin{equation}
\nabla^{2}=\frac{\partial^{2}}{\partial r^{2}}+\frac{1}{r}\frac{\partial}{\partial r}+\frac{1}{r^{2}}\frac{\partial^{2}}{\partial\phi^{2}}+\frac{\partial^{2}}{\partial z^{2}}.\label{eq:p2}
\end{equation}
Now we solve $\psi_{1}$ from  Eq.~(\ref{eq:qq2}). $\psi_{1}$ can
be chosen as
\begin{equation}
\psi_{1}=\left(\begin{array}{c}
f(r)e^{i(j-\frac{1}{2})\phi}\\
g(r)e^{i(j+\frac{1}{2})\phi}
\end{array}\right)e^{izp_{z}},\label{eq:a11}
\end{equation}
which is the common eigenstate of $\hat{p}_{z}$ and $-i\partial_{\phi}+\frac{1}{2}\sigma_{z}$
with eigenvalues $p_{z}$ and $j$. Plugging Eq.~(\ref{eq:a11}) into Eq.~(\ref{eq:qq2}) gives
\begin{eqnarray}
\left[\frac{d^{2}}{dr^{2}}+\frac{1}{r}\frac{d}{dr}+\left(E^{2}-M^{2}-p_{z}^{2}-\frac{(j-\frac{1}{2})^{2}}{r^{2}}\right)\right]f(r) & = & 0,\label{eq:qq4}\\
\left[\frac{d^{2}}{dr^{2}}+\frac{1}{r}\frac{d}{dr}+\left(E^{2}-M^{2}-p_{z}^{2}-\frac{(j+\frac{1}{2})^{2}}{r^{2}}\right)\right]g(r) & = & 0,\label{eq:qq5}
\end{eqnarray}
which are the Bessel equations of order $(j\mp\frac{1}{2})$. The
boundary conditions of $\psi_{1}$ at $r=0$ and $r=\infty$ require
that $E^{2}>M^{2}+p_{z}^{2}$. We can introduce a transverse momentum
$\alpha=\sqrt{E^{2}-M^{2}-p_{z}^{2}}$, then the eigen-energy becomes
$E=\lambda\sqrt{M^{2}+p_{z}^{2}+\alpha^{2}}$, with $\lambda=\pm1$
corresponding to the positive and negative modes. Now one can obtain
$\psi_{1}$ as
\begin{equation}
\psi_{1}=\left(\begin{array}{c}
J_{j-\frac{1}{2}}(\alpha r)e^{i(j-\frac{1}{2})\phi}\\
AJ_{j+\frac{1}{2}}(\alpha r)e^{i(j+\frac{1}{2})\phi}
\end{array}\right)e^{izp_{z}},\label{eq:p3}
\end{equation}
where $A$ is a constant to be determined. Since $\psi$ is also an
eigenstate of $-i\boldsymbol{\Sigma}\cdot\nabla$, then
\begin{equation}
-i\boldsymbol{\Sigma}\cdot\nabla\psi=s\epsilon\psi\label{eq:qq6}
\end{equation}
where $\epsilon=\sqrt{\alpha^{2}+p_{z}^{2}}$ is the magnitude of
the total momentum and $s=\pm1$ correspond to the two opposite helicities.
From Eq.~(\ref{eq:qq6}), one can get $-i\boldsymbol{\sigma}\cdot\nabla\psi_{1}=s\epsilon\psi_{1}$,
which leads to $A=\frac{i}{\alpha}(s\epsilon-p_{z})$ and
\begin{equation}
\psi_{2}=\frac{1}{M}(E+s\epsilon)\psi_{1}.\label{eq:p4}
\end{equation}
Finally, we obtain the eigenfunctions and corresponding eigen-energy
as follows,
\begin{equation}
\Psi_{\epsilon p_{z}js}^{(\lambda)}(t,r,\phi,z)=\frac{1}{4\pi\sqrt{X}}e^{-it\lambda\sqrt{X}+izp_{z}}\left(\begin{array}{c}
\sqrt{(X-\lambda s\epsilon)(\epsilon+sp_{z})}J_{j-\frac{1}{2}}(\alpha r)e^{i(j-\frac{1}{2})\phi}\\
is\sqrt{(X-\lambda s\epsilon)(\epsilon-sp_{z})}J_{j+\frac{1}{2}}(\alpha r)e^{i(j+\frac{1}{2})\phi}\\
\lambda\sqrt{(X+\lambda s\epsilon)(\epsilon+sp_{z})}J_{j-\frac{1}{2}}(\alpha r)e^{i(j-\frac{1}{2})\phi}\\
i\lambda s\sqrt{(X+\lambda s\epsilon)(\epsilon-sp_{z})}J_{j+\frac{1}{2}}(\alpha r)e^{i(j+\frac{1}{2})\phi}
\end{array}\right),\label{eq:p5}
\end{equation}
\begin{equation}
E_{\epsilon p_{z}js}^{(\lambda)}=\lambda\sqrt{M^{2}+\epsilon^{2}},\label{eq:p6}
\end{equation}
where $X=\sqrt{M^{2}+\epsilon^{2}}$, and $\lambda=\pm1$ correspond
to the positive and negative modes. All eigenfunctions are orthonormal,
\begin{equation}
\int dV\Psi_{\epsilon^{\prime}p_{z}^{\prime}j^{\prime}s^{\prime}}^{(\lambda^{\prime})\dagger}\Psi_{\epsilon p_{z}js}^{(\lambda)}=\delta_{\lambda^{\prime}\lambda}\delta_{j^{\prime}j}\delta_{s^{\prime}s}\delta(\epsilon^{\prime}-\epsilon)\delta(p_{z}^{\prime}-p_{z}).\label{eq:p7}
\end{equation}
\subsection{Axial vector current of a uniformly rotating system of massive Dirac fermions}
The Dirac equation in a uniformly rotating system with angular velocity
$\boldsymbol{\omega}=\omega\boldsymbol{e}_{z}$ can be written as \citep{Ambrus:2014uqa,Chen:2019tcp}
\begin{equation}
i\frac{\partial}{\partial t}\Psi(t,\boldsymbol{r})=\left(-i\gamma^{0}\boldsymbol{\gamma}\cdot\nabla+\gamma^{0}M-\omega\hat{J}_{z}\right)\Psi(t,\boldsymbol{r}).\label{eq:qq7}
\end{equation}
Compared with the free case in Sec.~\ref{subsec:free}, one can see
that, the eigenfunctions of Eq.~(\ref{eq:qq7}) are the same as the
free case, but with an energy shift $\Delta E=-j\omega$.
Now we consider a uniformly rotating system of massive Dirac fermions
with angular velocity $\boldsymbol{\omega}=\omega\boldsymbol{e}_{z}$,
where the interaction among fermions is ignored. This system is in
equilibrium with a reservoir, which keeps a constant temperature $T$
and constant chemical potential $\mu$. In the following, we will
calculate the axial vector current $J_{A}^{\mu}$ of this system.
According to the rotational symmetry along the $z$-axis of the system,
we can obtain $J_{A}^{x}=J_{A}^{y}=0$. Due to the absence of axial
chemical potential $\mu_{5}$ in our formalism, $J_{A}^{0}$ vanishes
\citep{Gao:2012ix}. The unique non-zero component is $J_{A}^{z}$.
From the approach of statistical mechanics used in Refs. \citep{Vilenkin:1978hb,Vilenkin:1979ui},
one can obtain
\begin{equation}
J_{A}^{z}=\sum_{\lambda,j,s}\int_{0}^{\infty}d\epsilon\int_{-\epsilon}^{\epsilon}dp_{z}\frac{\lambda}{e^{\beta\left[\sqrt{M^{2}+\epsilon^{2}}-\lambda(j\omega+\mu)\right]}+1}\Psi_{\epsilon p_{z}js}^{(\lambda)\dagger}\Sigma_{z}\Psi_{\epsilon p_{z}js}^{(\lambda)},\label{eq:qq8}
\end{equation}
where the Fermi-Dirac distribution has been inserted, and $\beta=1/T$.
Making use of the following series for Bessel function $J_{n}(x)$ with
$n\in\mathbb{N}$,
\begin{equation}
\left[J_{n}(x)\right]^{2}=\sum_{i=0}^{\infty}\frac{(-1)^{i}(2n+2i)!}{i![(n+i)!]^{2}(2n+i)!2^{2n+2i}}x^{2n+2i},\label{eq:p8}
\end{equation}
 Eq.~(\ref{eq:qq8}) becomes
\begin{equation}
J_{A}^{z}=\frac{T^{3}}{\pi^{2}}\sum_{N=0}^{\infty}\frac{\rho^{2N}}{2N+1}\sum_{n=0}^{\infty}C_{N,n}\frac{\Omega^{2n+1}}{(2n+1)!}\frac{d^{2n+1}}{d\alpha^{2n+1}}\mathrm{I}_{N}(\alpha,c),\label{eq:qq9}
\end{equation}
where we have defined four dimensionless quantities, $\rho=rT$, $\Omega=\omega/T$,
$\alpha=\mu/T$, $c=M/T$, and $C_{N,n}$, $\mathrm{I}_{N}(\alpha,c)$
are defined as
\begin{equation}
C_{N,n}=\sum_{j=0}^{N}\frac{(-1)^{N-j}}{(N-j)!(N+j)!(1+\delta_{j,0})}\left[\left(j+\frac{1}{2}\right)^{2n+1}-\left(j-\frac{1}{2}\right)^{2n+1}\right],\label{eq:qq10}
\end{equation}
\begin{equation}
\mathrm{I}_{N}(\alpha,c)=\int_{0}^{\infty}dyy^{2N+2}\left(\frac{1}{e^{\sqrt{y^{2}+c^{2}}-\alpha}+1}-\frac{1}{e^{\sqrt{y^{2}+c^{2}}+\alpha}+1}\right).\label{eq:qq12}
\end{equation}
The coefficient $C_{N,n}$ can also be expressed as follows
\begin{eqnarray}
C_{N,n} & = & \frac{1}{(2N)!}\left(x\frac{d}{dx}\right)^{2n+1}\left[x^{-N+\frac{1}{2}}(x-1)^{2N}\right]\bigg|_{x=1}\nonumber \\
 & = & \frac{2^{2N-2n-1}}{(2N)!}\left(\frac{d^{2n+1}}{dt^{2n+1}}\sinh^{2N+1}t\right)_{t=0},\label{eq:qq16}
\end{eqnarray}
where we have used the variable transformation $x=e^{t}$ in the second
line. According to the Taylor expansion of $\sinh t$, one can readily show
 $C_{N,n}=0$ for $n<N$. In principle, one can calculate $C_{N,n}$
for any $n\geqslant N$ from Eq.~(\ref{eq:qq16}). For example, for
$n=N,N+1$, one can obtain
\begin{eqnarray}
C_{N,N} & = & \frac{1}{2}(2N+1),\nonumber \\
C_{N,N+1} & = & =\frac{1}{24}(2N+1)^{2}(N+1)(2N+3).\label{eq:p9}
\end{eqnarray}
According to the calculation method in the appendixes of the recent
articles by some of us \citep{Zhang:2020ben,Fang:2021ndj}, the integral
$\mathrm{I}_{N}(\alpha,c)$ in Eq.~(\ref{eq:qq12}) can be expanded
at $c=0$ as follows,
\begin{equation}
\mathrm{I}_{N}(\alpha,c)=\sum_{l=0}^{\infty}\frac{(2l-2N-5)!!}{(2N+3)(-2N-5)!!(2l)!!}c^{2l}D_{N,l}(\alpha),\label{eq:q13}
\end{equation}
with $D_{N,l}(\alpha)$ expanded at $\alpha=0$ as
\begin{equation}
D_{N,l}(\alpha)=\sum_{k=0}^{\infty}(-1)^{l+k+N}\left(2-2^{2+2N-2l-2k}\right)\frac{(2l+2k-2N-2)!}{(2l-2N-4)!(2k+1)!}\frac{\zeta(2l+2k-2N-1)}{\pi^{2l+2k-2N-2}}\alpha^{2k+1}.\label{eq:q14}
\end{equation}
Plugging Eqs.~(\ref{eq:q13},~\ref{eq:q14}) into Eq.~(\ref{eq:qq9}),
one can obtain the series expansion of $J_{A}^{z}$ at $\rho=0$,
$\Omega=0$, $\alpha=0$, $c=0$ or $r=0$, $\omega=0$, $M=0$, $\mu=0$
as follows,
\begin{eqnarray}
J_{A}^{z} & = & T^{3}\sum_{N=0}^{\infty}\frac{\rho^{2N}}{(-2N-5)!!(2N+1)(2N+3)}\sum_{n=N}^{\infty}C_{N,n}\frac{\Omega^{2n+1}}{(2n+1)!}\sum_{l=0}^{\infty}\frac{(2l-2N-5)!!}{(2l)!!(2l-2N-4)!}c^{2l}\nonumber \\
 &  & \times\sum_{k=n}^{\infty}(-1)^{l+k+N}\left(2-2^{2+2N-2l-2k}\right)\frac{(2l+2k-2N-2)!}{(2k-2n)!}\frac{\zeta(2l+2k-2N-1)}{\pi^{2l+2k-2N}}\alpha^{2k-2n}.\nonumber \\
\label{eq:5}
\end{eqnarray}
If we only keep the linear term of $\Omega$ and set $\alpha=0$ in Eq.~(\ref{eq:5}), then
\begin{equation}
J_{A}^{z}=T^{2}\omega\sum_{l=0}^{\infty}(-1)^{l}\left(2-2^{2-2l}\right)(l-1)\zeta(2l-1)\frac{(2l-3)!!}{(2l)!!}\frac{c^{2l}}{\pi^{2l}}.\label{eq:p10}
\end{equation}
For the massless fermion case, we can get an analytic expression for $J_{A}^{z}$,
\begin{equation}
J_{A}^{z}=\left(\frac{T^{2}}{6}+\frac{\mu^{2}}{2\pi^{2}}\right)\frac{\omega}{(1-r^{2}\omega^{2})^{2}}+\frac{\omega^{3}(1+3r^{2}\omega^{2})}{24\pi^{2}(1-r^{2}\omega^{2})^{3}},\label{eq:p11}
\end{equation}
which is divergent as the speed-of-light surface is approached \citep{Ambrus:2014uqa}.

\section{Kubo Formula via Dimensional regularization}
\label{Dimensional}
\begin{figure}[ht]
  \centering
\includegraphics[width=3.12in,height=1.79in]{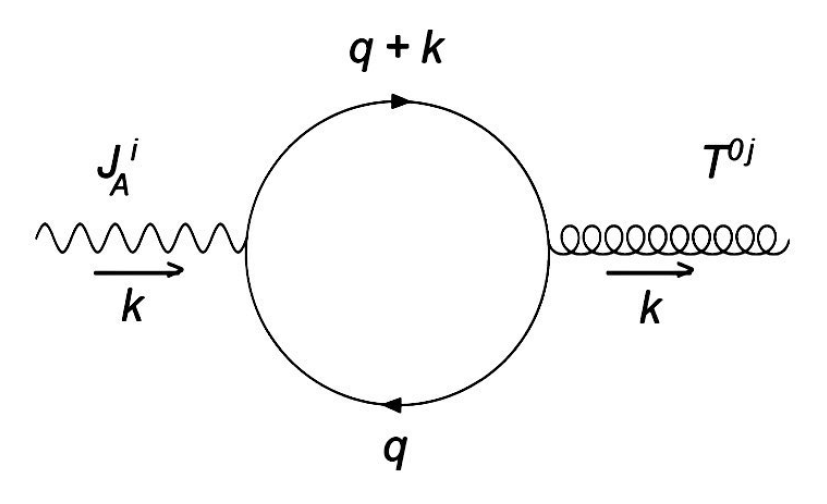}\\
  \caption{One-loop correction to the vortical conductivity~\cite{Landsteiner:2011cp}}              \label{fi:1}
\end{figure}
The Kubo formula relates the axial vorticalconductivity to the static Fourier component of the correlation between axial vector current $J_{A}^{i}$ and the stress tensor $T^{0j}$ via
\begin{equation}
J_{A}^{i}T^{0j}=2i\epsilon_{ijn}k_{n}\sigma ,
 \label{eq:v1}
\end{equation}
in the limit $\boldsymbol{k}\to0$. Ignoring interactions, LHS is represented by the one-loop thermal diagram in Fig.~\ref{fi:1}.~\citep{Landsteiner:2011cp,Hou:2012xg,Lin:2018aon,Amado:2011zx}. Calculating the thermal diagram with Matsubara formulation, we have
\begin{equation}\begin{aligned}
J_{A}^{i}T^{0j} \cong& \frac{i}{\beta}\sum_{v_{n}}\mathrm{tr}\frac{(q^{\mu}\gamma_{\mu}+M)\gamma_{i}\gamma_{5}[(q+k)^{\nu}\gamma_{\nu}+M][\gamma_{0}(q+\frac{k}{2})_{j}+\gamma_{j}q_{0}]}{\left(q^{2}-M^{2}\right)^{2}\left[(q+k)^{2}-M^{2}\right]^{2}} \\
=&	\frac{4i\epsilon_{ijn}k_{n}}{\beta}\sum_{v_{n}}\int\frac{d^{3}\boldsymbol{q}}{(2\pi)^{3}}\frac{\frac{1}{3}\boldsymbol{q}^{2}-v_{n}^{2}}{\left(\boldsymbol{q}^{2}+v_{n}^{2}+M^{2}\right)^{2}}\\
=&	\frac{2i\epsilon_{ijn}k_{n}}{\pi^{2}}\int dq \boldsymbol{q}^{2}\frac{1}{\beta}\sum_{v_{n}}\frac{\frac{1}{3}\boldsymbol{q}^{2}-v_{n}^{2}}{\left(\boldsymbol{q}^{2}+v_{n}^{2}+M^{2}\right)^{2}}  ,
\end{aligned}           \label{eq:r0}
\end{equation}
where the Matsubara frequency $v_{n}=(2n+1)\pi T-i\mu$. The summation and integral in  Eq.~(\ref{eq:r0}) appear UV-divergent and we apply the dimensional regularization by extending the spatial components of the loop momentum from 3-dimensional to $D$-dimensional, i.e.
\begin{equation}
\int\frac{d^3\boldsymbol{q}}{(2\pi)^3}\rightarrow\int\frac{d^D\boldsymbol{q}}{(2\pi)^D}  .
\end{equation}
It is straightforward to obtain that
\begin{equation}
\begin{aligned}
\sigma  =&-\frac{\omega_{D}}{(2\pi)^{D}}\sum_{v_{n}}\left[\frac{1}{D}B\left(1+\frac{D}{2},1-\frac{D}{2}\right)\left(v_{n}^{2}+M^{2}\right)^{\frac{D}{2}-1} \right.\\
&\left.-B\left(\frac{D}{2},2-\frac{D}{2}\right)v_{n}^{2}\left(v_{n}^{2}+M^{2}\right)^{\frac{D}{2}-2}\right]\\
 = &-\frac{\pi\omega_{D}T}{(2\pi)^{D}\sin\frac{\pi D}{2}}\sum_{n=0}^{\infty}v_{n}^{D-2}\left[\left(1+\frac{M^{2}}{v_{n}^{2}}\right)^{\frac{D}{2}-1} -(2-D)\left(1+\frac{M^{2}}{v_{n}^{2}}\right)^{\frac{D}{2}-2}\right]\\
 = &\sigma|_{M=0}+\Delta\sigma ,
\end{aligned}      \label{eq:r1}
\end{equation}
where, the D-dimensional solid angle $\omega_{D}=\frac{2\pi^{D}}{\Gamma(\frac{D}{2})}$, and $B\left(x,y\right)$ is the beta function. The last line of Eq.~(\ref{eq:r1}) separate $\sigma$ into two terms, where $\sigma|_{M=0}$ is the vortical conductivity of massless fermions, and $\Delta\sigma$ is the mass correction of the vortical conductivity. Here,
\begin{equation}
\sigma|_{M=0}=-\frac{\pi\omega_{D}T}{(2\pi)^{D}\sin\frac{\pi D}{2}}\sum_{n=0}^{\infty}v_{n}^{D-2}(D-1)  ,           \label{eq:r3}
\end{equation}
and
\begin{equation}\begin{aligned}
\Delta\sigma=&-\frac{\pi\omega_{D}T}{(2\pi)^{D}\sin\frac{\pi D}{2}}\sum_{n=0}^{\infty}v_{n}^{D-2}\left[\left(1+\frac{M^{2}}{v_{n}^{2}}\right)^{\frac{D}{2}-1}-(2-D)\left(1+\frac{M^{2}}{v_{n}^{2}}\right)^{\frac{D}{2}-2}-D+1\right].     \end{aligned} \label{eq:r4}
\end{equation}
Let the dimensionality $D=3-\epsilon$ and taking the limit $\epsilon\to 0$, we find
\begin{equation}
\begin{aligned}
\sigma|_{M=0}	=& 	-\frac{\omega_{D}(D-1)T^{D-1}}{(2\pi)^{D-1}\sin\frac{\pi D}{2}}\mathrm{Re}\zeta\left(2-D,\frac{1}{2}-i\frac{\mu}{2\pi T}\right)  \\
\to& \frac{T^{2}}{6}+\frac{\mu^{2}}{2\pi^{2}} (\epsilon\to 0).
\end{aligned}             \label{eq:r14}
\end{equation}
Upon expanding $\Delta\sigma$ in the power of $M^2$, the leading term, the $M^2$ term, is of the form $0\times\infty$ at $D=3$ and the limit has to be taken carefully. Let $c_{D}$ be the coefficient of $M^{2}$, we have
\begin{equation}
\begin{aligned}
c_{D}=&	-\frac{\pi\omega_{D}T}{(2\pi)^{D}\sin\frac{\pi D}{2}}(\frac{D}{2}-1)(D-3)\mathrm{Re}\sum_{n=0}^{\infty}v_{n}^{D-4}  \\
\to&-\left(-\frac{T}{\pi}\right)\frac{1}{2}(1-\epsilon)(-\epsilon)(\pi T)^{-1}(1-2^{-1})\mathrm{Re}\zeta\left(1+\epsilon,\frac{1}{2}-i\frac{\mu}{2\pi T}\right)\to-\frac{1}{4\pi^{2}}  .
\end{aligned}               \label{eq:r6}
\end{equation}
For the higher power in $M^2$, however, the naive limit works. Taken together, we end up with the limit
\begin{equation}
\begin{aligned}
\Delta\sigma=-\frac{M^{2}}{4\pi^{2}}+\frac{T}{2\pi}\sum_{n=0}^{\infty}v_{n}\left[\left(1+\frac{M^{2}}{v_{n}^{2}}\right)^{\frac{1}{2}}+\left(1+\frac{M^{2}}{v_{n}^{2}}\right)^{-\frac{1}{2}}-2\right] .
\end{aligned}         \label{eq:r8}
\end{equation}
Adding Eq.~(\ref{eq:r14}) and Eq.~(\ref{eq:r8}), we replicate Eq.~(\ref{eq:sigma}).  We have also verified that the same result emerges with the Pauli-Villars regularization.

\bibliographystyle{unsrt}
\bibliography{ACVE}
\end{document}